\documentclass[onecolumn]{svjour3}          % onecolumn

\usepackage{seqsplit}
\usepackage[hyphens]{url}
\usepackage[ruled]{algorithm2e}
\usepackage[T1]{fontenc}
\usepackage{times}
\usepackage{amsmath}
\usepackage{booktabs,caption}
\usepackage[flushleft]{threeparttable}
\usepackage{float}
\usepackage{algpseudocode}
\usepackage{algorithm2e}
\usepackage{mathtools}
\usepackage{multirow}
\usepackage{graphicx}
\usepackage{slashbox,multirow}
\usepackage{arydshln}
\usepackage{xcolor}
\usepackage{breqn}
\usepackage{balance}
\DeclarePairedDelimiter\abs{\lvert}{\rvert}%

\usepackage{amssymb}% http://ctan.org/pkg/amssymb
\usepackage{pifont}% http://ctan.org/pkg/pifont

\newcommand{\cmark}{\ding{52}}%
\newcommand{\xmark}{\ding{54}}%

\def\BibTeX{{\rm B\kern-.05em{\sc i\kern-.025em b}\kern-.08em
    T\kern-.1667em\lower.7ex\hbox{E}\kern-.125emX}}

\makeatletter
\def\endthebibliography{%
  \def\@noitemerr{\@latex@warning{Empty `thebibliography' environment}}%
  \endlist
}
\usepackage[T1]{fontenc}
\usepackage{float}
\usepackage{amsmath}
\usepackage{graphicx}
\usepackage{setspace}

\floatstyle{ruled}
\newfloat{algorithm}{tbp}{loa}
\providecommand{\algorithmname}{Algorithm}
\floatname{algorithm}{\protect\algorithmname}

\usepackage[caption=false,font=footnotesize]{subfig}
\usepackage{balance}

\smartqed  % flush right qed marks, e.g. at end of proof
\usepackage{graphicx}
\usepackage{latexsym}
\begin{document}

\title{Novel Randomized Placement For FPGA Based Robust ROPUF with Improved Uniqueness
}
\subtitle{}

%\titlerunning{Short form of title}        % if too long for running head

\author{Arjun Singh Chauhan         \and
        Vineet Sahula \and
        Atanendu Sekhar Mandal
}

%\authorrunning{Short form of author list} % if too long for running head

\institute{Malaviya National Institute of Technology \at
              JLN Marg, Jaipur, Rajasthan, India \\
              Tel.: +91-9785324861\\
             \email{2015rec9055@mnit.ac.in}           %  \\
        \and
              Malaviya National Institute of Technology \at
              JLN Marg, Jaipur, Rajasthan, India \\
              Tel.: 0141-2713336\\
              {\footnotesize{The final online version of this pre-print can be found at https://doi.org/
10.1007/s10836-019-05829-5}}\\
             \email{sahula@ieee.org}           %  \\
		\and
			 Cognitive Computing Group, CEERI \at
             Pilani, Rajasthan, India \\
            \email{atanu@ceeri.ernet.in}           %  \\
}

\date{This is pre-print of published paper in Springer's Journal of Electronic Testing, Dec. 2019}
% The correct dates will be entered by the editor

\maketitle

\abstract
{
The physical unclonable functions (PUF) are used to provide software as well as hardware security for the cyber-physical systems. They have been used for performing significant cryptography tasks such as generating keys, device authentication, securing against IP piracy, and to produce the root of trust as well. However, they lack in reliability metric. We present a novel approach for improving the reliability as well as the uniqueness of the field programmable gated arrays (FPGAs) based ring oscillator PUF and derive a random number, consuming very small area ($ < 1\%$) concerning look-up tables (LUTs). We use frequency profiling method for distributing frequency variations in ring oscillators (RO), spatially placed all across the FPGA floor. We are able to spot suitable locations for RO mapping, which leads to enhanced ROPUF reliability. We have evaluated the proposed methodology on \textit{Xilinx -7} series FPGAs and tested the robustness against environmental variations, e.g. temperature and supply voltage variations, simultaneously. The proposed approach achieves significant improvement (i) in uniqueness value upto $49.90\%$, within 0.1\% of the theoretical value (ii) in the reliability value upto $99.70\%$, which signifies that less than 1 bit flipping has been observed on average, and (iii) in randomness, signified by passing NIST test suite. The response generated through the ROPUF passes all the applicable relevant tests of NIST uniformity statistical test suite.
}

\vspace{-1em}
\section{Introduction}

Nowadays, information security, integrity and privacy are the primary attention for each software and hardware subsystems, since the compromised subsystems introduce vulnerability by getting exploited due to various loopholes. Most of these hardware subsystems have been designed by appropriating application specific integrated circuits (ASIC) and field programmable gated arrays (FPGA). FPGA based security solutions are distinguished due to reconfigurable architectures. To secure the information, cryptographic primitives have been used to advance an essential level of information security and integrity. The primary requirement for the cryptographic primitive is to protect the secret keys against various security threats, i.e. key stealing, random guessing, etc. The earlier approaches to protect secret keys against random guessing attacks attempt to increase the entropy of secret key and keep it high enough to pass randomness benchmarks. These random keys can be stored in secure non-volatile memory (NVM) of the hardware device, and appropriated during cryptographic operations.

Nevertheless, these keys can not resist against physical attacks, i.e. side channel parameter extraction, device tempering, etc \cite{Side_Channel_Strong_PUF} \cite{Fault_Injection_Attacks}. These attacks can extract at least partial information of the secret key and predict the remnant. However, hardware-enabled random number generation (RNG) methods provide a suitable alternative, instead of storing it. Moreover, physical unclonable function (PUF) has been utilized to produce random information from the hardware, unlike NVM based keys.

The physical unclonable function has been used to generate a random unique identification sequence by employing stochastic process variability of the silicon-based chips. An electronic circuit is used to produce a $n$ bit binary unique identification sequence, each from a device. These identification sequences are appropriated to provide an essential level of hardware security against IC overbuilding, IP piracy and are also used to protect against device tempering in hardware, i.e. ASIC and FPGA \cite{Hardware_Primer} \cite{Securing_IoT_R_Karri}; moreover, the same can also be used for hardware-based cryptographic keys generation  \cite{Reference_ROPUF_Srinidevdas}, hardware device authentication \cite{S_Devadas_First} \cite{Pappu2001} and hardware-based random number generation \cite{hardware_RNG}. Numerous methods have been proposed to produce hardware based random number generations, i.e. linear feedback shift registers \cite{Generalized_LFSR}, power supply based RNG \cite{Power_Source_Noise_Based_RNG}, ring oscillator based RNG \cite{RO_Based_TRNG}, etc.

FPGA based security solutions are quite prominent and widely adopted due to reconfigurability. Therefore, light-weight and secure cryptographic algorithms have been implemented on FPGA, and employed for the IoT security \cite{IITKGP_Attack_IoT}. 

The work related to unclonable function has been started with physical one-way function. The one-way function has been used to produce random optical patterns by appropriating the optical characteristics of a material. These optical characteristics have an approximately zero probability for regeneration, without knowing the exact angle information \cite{Pappu2001}. The silicon-based physical unclonable function has been introduced to produce random information by employing a digital circuit, which is termed as arbiter based PUF design \cite{S_Devadas_First} \cite{S_Devadas_Extractkeys}. The arbiter PUF design performs a delay comparison between two symmetrically routed path, where each path consists of the equal number of logic gates. The design exploits the random manufacturing variations and derives a random signature from it. There are many more PUF designs have been proposed to produce random binary sequences, i.e. ring oscillator \cite{Reference_ROPUF_Srinidevdas}, latch and flip-flops enabled \cite{Buskeeper_PUF} \cite{RS_Latch_PUF} \cite{Review_Butterfly_PUF}, static random access memory (SRAM) \cite{SRAM_Based_PUF}, etc. 

PUF can be categorized in mainly two types on the basis of challenge-response pairs (CRP); 1) Strong, and 2) Weak. A weak PUF typically consists of less number of CRP. However, the strong PUF has many number of CRP. The typical application of the strong PUF is the device authentication, where a large number of CRP requires. The example of strong PUF is arbiter based, optical based, etc. The weak PUF typically used for cryptographic key generation, i.e. ring oscillator PUF (ROPUF), Latch enabled, SRAM based, etc.

The ring oscillator based PUF has been designed by employing multiple ring oscillator, and each ring oscillator produces different frequency due to its manufacturing variations. These frequencies have been used to generate random unique identification sequences. ROPUF design is less affected by routing skew as compared to arbiter based design. {\color{blue} However, ring oscillators are more prone to uncontrolled environmental noises, i.e. voltage ($V$), temperature ($T$) and device aging ($\lambda$) \cite{TEMP_Variation_Review_1}. These noises revise the internal propagation delay of ring oscillators; consequently, the frequency variation appears to be significant, which causes unreliable response bit generation or response bit flipping. }

The security of the cryptographic algorithms is dependent on the secure key, and the keys should be reliable and reproducible at different environmental conditions. Therefore unreliable cryptographic key can corrupt the entire message after encryption, which is hard to decrypt at the other end, e.g. secure hash algorithm (SHA) can change 50\% of message bits with a single bit flipping in secret key. Similarly, high bit flipping rate can increase the false acceptance rate (FAR) and false rejection rate (FRR) during device authentication application. Therefore reliable response generation through PUF is a relevant and essential problem to be solved. However, the uniqueness enhancement can reduce the FAR and FRR, although randomness can improve the unpredictability of the secure key.

\section{Related Work}

Author Devadas et al. have started extensive work related to ring oscillator PUF \cite{Reference_ROPUF_Srinidevdas}. After that, various researchers have proposed numerous numbers of research articles to address and solve the ring oscillator PUF related problems. The response produced by the ROPUF deficits with unreliable response bits. The increment in unreliable response bits leads to produce misidentification errors, therefore inter-PUF and intra-PUF hamming distances should keep as large as possible. There are many approaches has been proposed to improve the frequency response of the ring oscillator PUF and the improvement in frequency response can improve both distances. 

Devadas et al. have introduced the very first ROPUF reliable design, and the approach provides a significant improvement in the frequency response. The author improves the ring oscillator selection logic based on the frequency difference from a group of $k$ ring oscillators. A pair of ring oscillator has been selected out of $k$ ring oscillators to maximize the frequency difference. The approach provides a significant improvement in reliability, but the area utilization has been increased by $k$ times.

The other significant work relating to reliability improvement is the configurable ring oscillator (CRO), which was proposed by Maiti et al. \cite{Maiti_J_Crypto}. The author proposed a new design consists of configurable ring oscillators. These ring oscillators are configured in such a way that frequency difference between selected ring oscillator should be maximized. The approach improves the reliability at $90nm$ technology, and the reliability is $>99.2\%$ in the presence of environmental variations. There are other implementations of the CRO have been proposed to address the reliability and uniqueness metric issues \cite{Configurable_RO_1} \cite{Reconfigurable_ROs}.

There is a remarkable contribution has been made to correct bit error rate (BER) by introducing post-processing techniques, some of these are the error correction techniques, i.e. temporal majority voting (TMV), Bose-Chaudhuri- Hocquenghem (BCH) codes, etc \cite{SRAM_Error_Correction_BCH} \cite{Temporal_Voting}. There are other techniques such as fuzzy extraction, which has been used to provide strength to the device authentication in the presence of environmental noise \cite{SRAM_Error_Correction_Fuzzy_Extractor}. These approaches require extra resources on the hardware and also requires spare computation time for processing.

The FPGA based ROPUF design has limited access to reduce the effect of environmental noise, therefore some authors proposed LUT level modification to enhanced the RO-PUF performance. The author suggested a LUT based self compare structure, which can generate $256$ bit response by tuning the LUT delay lines \cite{Selfcompare}. The approach provides an improvement in uniqueness, but the fine-tuning of delay line originates unreliable response bits. Moreover, the author proposed an adaptive tuning circuit, which can reduce unreliable response bits, but the tuning design itself requires extra hardware on FPGA. The author Wei Yan et al. proposed an approach based on phase calibrated ring oscillators, the phase calibration technique has been employed to measure fast and accurately RO frequencies. The response generated through phase calibrated ROPUF design has passed all NIST statistical tests along with very less unreliable bits $< 1\%$ in the presence of the temporal environmental variations \cite{Fatema_Teh_NIST_Stat}.

The ROPUF reliability enhancement has been accomplished by increasing the frequency difference. Apart from the CRO and 1-out-of-k approach, there are numerous methods have been proposed to improve the ROPUF reliability. These approaches extract the ring oscillator frequency variation using on-chip frequency monitors and choose a pair of ring oscillators, which provides a significantly large frequency difference. The logic for ring oscillator selection and the frequency difference enhancement are different for each approach. The authors have proposed methods related to temperature awareness frequency measurement \cite{Systematic_2_TAC} \cite{Frequency_Offset}. These approaches prefer a ring oscillator pair which has better temperature stability. Moreover, the reliability improvement is significant, but the approach is limited to temperature variations, and it also requires some mechanism to perform temperature characterization of ring oscillators. {\color{blue} The authors F. Kod\'{y}tek et al. present an approach to improve the randomness characteristics using the entropy of the counter bit position, while counter is subjected to the ring oscillator frequency. The method possess significant randomness and passes NIST statistical methods, however the approach shows high bit flipping in the presence of voltage variation {\color{blue}\cite{NIST_Reviewer_1}}. }

The other works are related to grouping based approach, where a group or specific order has chosen with several frequencies, and these frequencies are selected in such a manner that the frequency difference has been improved \cite{DP_Based_Group} \cite{Group_Based}.  The author defines a threshold frequency ($f_{th}$), which is the safe limit for frequency difference and the approach search for the ring oscillators, which provides at least $f_{th}$ frequency difference. The selection of $f_{th}$ is uncertain, and it depends on the characterization accuracy, environmental variations and device family.

{\color{blue} The characterization phase for all of these approaches is almost the same, except the accuracy and device type. The accuracy and computation time for the characterization phase relies on the counter enable duration, system clock frequency, number of inverters in ring oscillators, and routing propagation delay. Some of the quite prominent frequency monitoring methods have been proposed for accurate frequency measurement \cite{RO_Characterization_Maiti} \cite{RO_Characterization_Host} \cite{Characterization_Reviewer_1}. These methods have been implemented for different device families.

 The author Chauhan et al. presented an approach, which improves the reliability of ROPUF design by increasing frequency difference among ring oscillators \cite{Arjun_IEEE_CONECCT}. The authors have proposed, the enhancement in frequency difference using a new characterization phase, biased placement and k-means clustering. The overall reliability improvement is apparent due to the frequency difference enhancement. However, the other parameters such as uniqueness and uniformity diminish, due to biased placement. Furthermore, the approach is limited to temperature variations. Later, the approach has been modified to improve the randomness and uniqueness along with reliability \cite{ARJUN_VLSID}. The authors additionally, employ the randomized placement to enhance the uniqueness, whereas randomness is improved using random frequency allocation with an LFSR design. The sequences produced by modified ROPUF design have passed NIST statistical tests for randomness. However, the reliability certainly get reduced for the extreme voltage conditions. Furthermore, the routing validation step increases ROPUF synthesis and implementation efforts, making it a comparatively time-consuming process.}

The rest of the paper is organized as follows. The problem motivation and current novel contribution are discussed in section \ref{sec:motivation} and \ref{sec:contribution_in_paper}, respectively. In section \ref{sec:Primitives} we have discussed ring oscillator design, modeling and effect of measurement uncertainties. The proposed approach and the experimental results are presented in sections \ref{sec:Proposed_Method} and \ref{sec:Experiment_results}, respectively. The manuscript has concluded in section \ref{sec:Conclusions}.

\vspace{5em}

\section{Motivation\label{sec:motivation}}
{
\color{blue}

The ring oscillator PUF is considered a weak PUF, since it can only produce a small number of challenge-response pairs. Therefore, it is widely used for cryptographic key generation, where cryptographic core obfuscates the CRP correlation. Some of the researchers have utilized weak PUF with strong PUF and pseudo random number generator (PRNG) to increase the obfuscation, and thus are able to improve the resistance against machine learning attacks \cite{Dual_Mode_PUF} \cite{Multi_PUF}. The PUFs are vulnerable to machine learning attacks, which are capable of replicating challenge-response correlation by employing intelligent machine learning (ML) approaches. Similar, to the ML attacks, a genetic algorithm (GA) based attack has been explored to predict the response of ROPUF design. These approaches can predict CRP with a prediction rate of upto $96\%$ \cite{GA_Based_Attack} \cite{GA_Based_RS}.

The authors proposed an improved version of the existing ROPUF design in order to make it as a  strong PUF. Further, the reliable PUF-ID generator is used to provide challenges instead of applying challenges directly to the PUF. Here, the reliable PUF-ID introduces obfuscation between input and output and thus reduces the CRP correlation. However, the environmental variations have not been considered with the approach \cite{PUFID_Generator_GU_ET_AL}. Similarly, the authors have proposed dual-mode PUF to improve the ML resistance \cite{Dual_Mode_PUF}, where the approach utilizes two modes; i) counting and ii) state stabilization. However, the obfuscation increases the sensitivity to environmental noise. Thus authors in \cite{Dual_Mode_PUF}, report a $>13\%$ bit flipping occurrence due to temperature variations. They do not consider the effect of supply voltage variations. As the supply voltage variation modifies ring oscillator pulse width nonlinearly, the effect of such variation is more significant as compared to effect due to temperature variation.
 
These attempts accommodate ways to secure device authentication against ML attacks by combining multiple PUFs, which could lead to increased sensitivity to environmental variations. Hence, a single but more reliable PUF is preferable in order to improve false rejection ratio (FRR) as well as false acceptance ratio (FAR). The other solution is to utilize the error correction schemes, albeit at the cost of FPGA chip area overhead, constraining how large a number of PUFs could be realized on the remaining area.
}

\subsection{Our Contribution \label{sec:contribution_in_paper}}

The work proposed in this paper is substantial extension of authors' work for ROPUF reliability and uniqueness enhancement using randomized and biased placement on FPGA \cite{ARJUN_VLSID} \cite{Arjun_IEEE_CONECCT}. Following are our additional contributions in this manuscript.

\begin{itemize}
\item The ring oscillator frequency difference is maximized using augmented clustering-based approach. We propose an improved K-means based grouping approach by novel approach for global maximum based inter-centroid frequency selection. This novel augmentation in algorithm results in improving the frequency difference characteristics of the ROPUF.

\item The further improvement in the minimum pairwise frequency difference has been obtained by employing novel scheme for centroid relocation based difference maximization. The proposed approach provides the enhancement in the frequency difference whenever the number of ring oscillators in ROPUF increase.

\item The randomness evaluation has been performed using NIST statistical tests. We have introduced a controllable randomness factor to configure the randomness of the proposed ROPUF, which also we have validated by employing NIST statistical tests and minimum entropy analysis.

\item We propose the elimination of routing validation scheme in design flow. Alternately, we effect modifications in hardware design, which altogether omits synthesis \& implementation and thus saves time/efforts.

{\color{blue}

\item We have included a detailed analysis of the improved k-means clustering, overall design time, minimum entropy. We have performed experimentation on increased the number of hardware (FPGA devices) to achieve higher confidence in the results.
}

\end{itemize}

\section{Ring Oscillator Physical Unclonable Functions\label{sec:Primitives}}

\begin{figure}[tbh]
\centering{}\includegraphics[width=1\columnwidth]{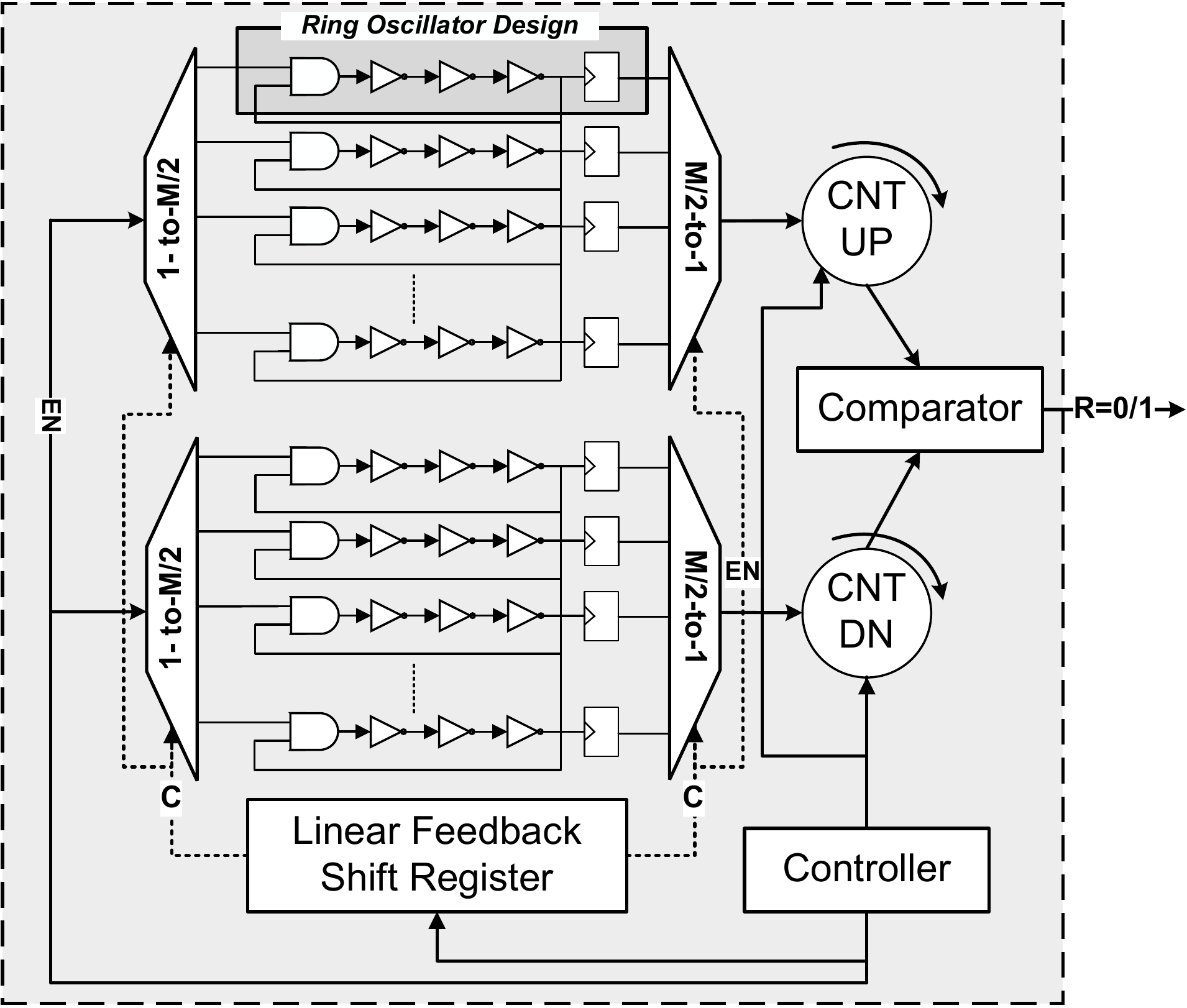}\caption{The ring oscillator based PUF design consists of M ring oscillators. The input and output can be represented as a tuple of challenge-response pair $(C, R)$. The challenges have applied using a maximal length LFSR design.
\label{fig:ROPUF_Design_JETTA}}
\end{figure}

The ring oscillator PUF design consists of $M$ number of ring oscillators divided into two groups as shown in Figure \ref{fig:ROPUF_Design_JETTA}, where each group consists of $\frac{M}{2}$ ring oscillators and these ring oscillators are placed over the entire FPGA chip. Presuming that manufacturing variation modifies the frequency of each ring oscillator. A pair of ring oscillator has been selected, each from a group using the input challenge ($C$), and finally pulse difference has been evaluated using a comparator design. The response $(R)$ is the binary output generated through the comparator design.
 
\subsection{Ring Oscillator Delay Modeling}

A typical ring oscillator has been designed with the odd number of inverter and enable logic element. The propagation delay $\tau_g$ of a logic element due to manufacturing variation can be assumed as the multivariate Gaussian distribution, and it is dependent on the process parameters, i.e. the threshold voltage ($V_{th}$), the channel length ($L$), width ($W$), etc.

The $\tau_g$ is composed of mainly three components; 1.) average delay component $\tau_{g}^{avg}$, 2.) systematic delay component $\tau_{g}^s$ and, 3.) random delay component  $\tau_{g}^r$. The propagation delay of $i^{th}$ logic element is the aggregation of all three delay components and it can be represented as (\ref{eq:ROPUG_Delay_eq_2}).

\begin{equation}
	\tau_{g_{i}}=\tau_{g_{i}}^{avg}+\tau_{g_{i}}^s+\tau_{g_{i}}^r
	\label{eq:ROPUG_Delay_eq_2}
\end{equation}

The propagation delay ($d$) of $n$ logic element based ring oscillator is manifested in (\ref{eq:ROPUG_Delay_eq_3}).

\begin{equation}
d=\left( \sum_{i=1}^{n} \left(\tau_{g_{i}}^{nom} + \tau_{g_{i}}^{s} + \tau_{g_{i}}^{r} \right)   \right) + d_{w}
\label{eq:ROPUG_Delay_eq_3}
\end{equation}

Here, $d_{w}$ is the routing path delay of the ring oscillator. The average value appears due to the effective value of process parameters. The systematic delay component arises due to non-uniformity during the fabrication, whereas the random delay component is present due to irregular doping concentration in transistors. Therefore each ring oscillator produces a distinct frequency and further it is used to generate random signatures.

\subsection{ROPUF response modeling}

The ring oscillator PUF design produces random numbers by performing comparison in between a pair of the ring oscillator. A pair of digital counters has been utilized to capture ring oscillator pulses ($\alpha$) for $t_{on}$ duration, therefore the propagation delay ($d$), and the RO frequency ($f$) at an instance is manifested in (\ref{eq:delay_equation}).

\begin{equation}
	d=\frac{t_{on}}{2\alpha} \text{ , } f=\frac{\alpha}{t_{on}} 	
	\label{eq:delay_equation}
\end{equation}

{\color{blue}
The delay difference between $i^{th}$ and $j^{th}$ ring oscillators can be evaluated using (\ref{eq:delay_equation_diff}).

\begin{equation}
	\Delta d= - \frac{t_{on}\Delta\alpha}{2\alpha_{i}\alpha_{j}} \text{ , } \Delta f=\frac{\Delta\alpha}{t_{on}} 	
	\label{eq:delay_equation_diff}
\end{equation}

From (\ref{eq:delay_equation_diff}), we can conclude that, $\Delta F \propto \Delta\alpha \propto \Delta d$.}
Therefore the ROPUF response can be evaluated in terms of either count difference $(\Delta\alpha)$ or frequency difference $(\Delta f)$, which is manifested in (\ref{eq:response_eq_1}).

\begin{equation}
 R= 
 \begin{cases}
    0,& \text{if } \Delta f \geq 0\\
    1,              & \text{otherwise}
\end{cases}
\label{eq:response_eq_1}
\end{equation}

A noise $(\epsilon)$ has been introduced during count value extraction from counter, it can be expressed using (\ref{eq:freq_diff_Eq}).

\begin{equation}
	\Delta f_{\epsilon} = \Delta f_{\epsilon_{m}}  + \Delta f_{\epsilon_{t}}   
	\label{eq:freq_diff_Eq}
\end{equation}

Here, $\Delta f_{\varepsilon_{t}}$ and $\Delta f_{\varepsilon_{m}}$ are the environmental noise and measurement noise. The error-free frequency difference ($\Delta\overline{f}$) can be expressed as the aggregation of systematic ($\Delta\overline{f}_{sys}$), random ($\Delta\overline{f}_{rand}$) and wire frequency difference ($\Delta\overline{f}_{w}$). Therefore the actual frequency difference ($\Delta f$) can be expressed as (\ref{eq:freq_diff_actual}).
\begin{equation}
	\Delta f = \Delta\overline{f}_{sys} + \Delta\overline{f}_{rand} + \Delta\overline{f}_{w} +\Delta f_{\varepsilon_{t}} + \Delta f_{\varepsilon_{m}}
	\label{eq:freq_diff_actual} 
\end{equation}

The unreliable response is referred as the frequency difference ($\Delta f$) sign change for same input during difference conditions.

\subsection{Effect of Measurement Error}
The frequency difference measurement error $\Delta f_{\epsilon_{m}}$ is random in nature and it can be modeled as $\mathcal{N}(0, \sigma_{\varepsilon_{m}}^2)$, here mean is assumed to be zero and $\sigma_{\varepsilon_{m}}$ is the standard deviation for measurement error. The maximum frequency difference measurement error $(\Delta f_{\epsilon_{m}}^{max}) = 6\sigma$ (range is $\pm3\sigma$) for 99.7\% confidence interval. Therefore highly deviated ring oscillators increase bit flipping possibility by increasing measurement error.

\subsection{Effect of Environmental Error}

The environmental variations modifies the ring oscillator frequency difference. The source of environmental variation is temperature (T), supply voltage (V) and device aging ($\lambda$). Rewriting (7) in (8).

\begin{equation}
\Delta f = \Delta\overline{f} + \Delta f_{\epsilon_{t}} + \Delta f_{\epsilon_{m}}
\end{equation}

From (8), it is clear that reliable condition (no frequency sign change) can be achieved by either (\ref{eq:stable_1}) or (\ref{eq:stable_2}).
\begin{equation}
{\color{blue}
	\abs{\Delta\overline{f}} > (\abs{\Delta f_{\varepsilon_{t}} + \Delta f_{\varepsilon_{m}}})}
	\label{eq:stable_1} 
\end{equation}
\begin{equation}
	\Delta\overline{f} * (\Delta f_{\varepsilon_{t}} + \Delta f_{\varepsilon_{m}}) > 0
	\label{eq:stable_2}
\end{equation}

The condition (\ref{eq:stable_1}) itself is a complete one, but if (\ref{eq:stable_1}) does not satisfy then (\ref{eq:stable_2}) can be checked for stability.

\section{Proposed Methodology \label{sec:Proposed_Method}}

The proposed approach has been used to increase the frequency difference ($\Delta f$) such that all the ring oscillator pairs should satisfy (\ref{eq:stable_1}), because the effect of noise ($\Delta f_{\epsilon}$) in frequency difference can be compensated when the frequency difference is comparatively large. Therefore highly separated ring oscillator frequencies are preferred to improve the reliability.

\begin{algorithm}[tbh]
\small
\SetAlgoLined
\SetKwFunction{findLocation}{findLocations}
\SetKwFunction{sortFunction}{sortFunction}
\SetKw{Phasezero}{Phase-0 : Constraint Creation} 
\SetKw{Phaseone}{Phase-1 : Frequency Variation Profiling} 
\SetKw{Phasetwo}{Phase-2 : ROPUF Creation } 
\SetKw{constgen}{\textbf{1} :Generate Place \& Route Constraint} 
\SetKw{synthImpl}{\textbf{2} :Capture Frequency Data using Monitor} 
\SetKw{stablero}{\textbf{1} : Erroneous Frequency Rejection} 
\SetKw{grouping}{\textbf{2} : Clustering Based Grouping} 
\SetKw{selection}{\textbf{3} : Group Frequency Selection} 
\SetKw{allocation}{\textbf{4} : Ring Oscillator Allocation} 
\SetKw{validation}{\textbf{5} : Routing Path Validation} 
\begin{raggedright}
\Phasezero\\
\Begin{
\small
1: Place \& Route Constraint Creation for Characterization \\
}
\Phaseone\\
\Begin{
\small
1: Capture Frequency Data using On-chip Frequency Monitors\\
2. Erroneous Frequency Rejection (Post Process)\\
}
\Phasetwo\\
\Begin{
	\small	
	1: Clustering Based Grouping.\\
	2: Centroid relocation for difference maximization.\\
	3: Ring oscillator group assignment. \\
	4: Randomized placement of ring oscillators.\\
	5: Create constraint files. 
}
\end{raggedright}
\caption{Work flow of the proposed approach}
\label{alg:Flow_of_the_approach} 
\end{algorithm}

{\color{blue}

The proposed approach is a two-phase approach, where the first phase extracts the process variation of the entire FPGA chip using $\mathcal{Z}$ ring oscillator instances. Each ring oscillator instance is placed to a single slice. During the second phase, we have selected M ring oscillator instances out of $\mathcal{Z}$ instances, in order to achieve a significant frequency difference, which improves the overall reliability, here $Z\gg M$. The manual placement \& routing constraints have been included to fix the placement and routing of ring oscillators. Since we have used fixed placement and routing, therefore no other logic can occupy the allocated resources.
}

The flow of the proposed approach is shown in Algorithm \ref{alg:Flow_of_the_approach}. All the constraints related to the first phase has been created during a pre-processing phase, which is termed as phase-0 or constraint creation phase.

\subsection{Phase-0 : Constraint Creation}

\begin{figure}[tbh]
\centering{}\includegraphics[width=0.9\columnwidth]{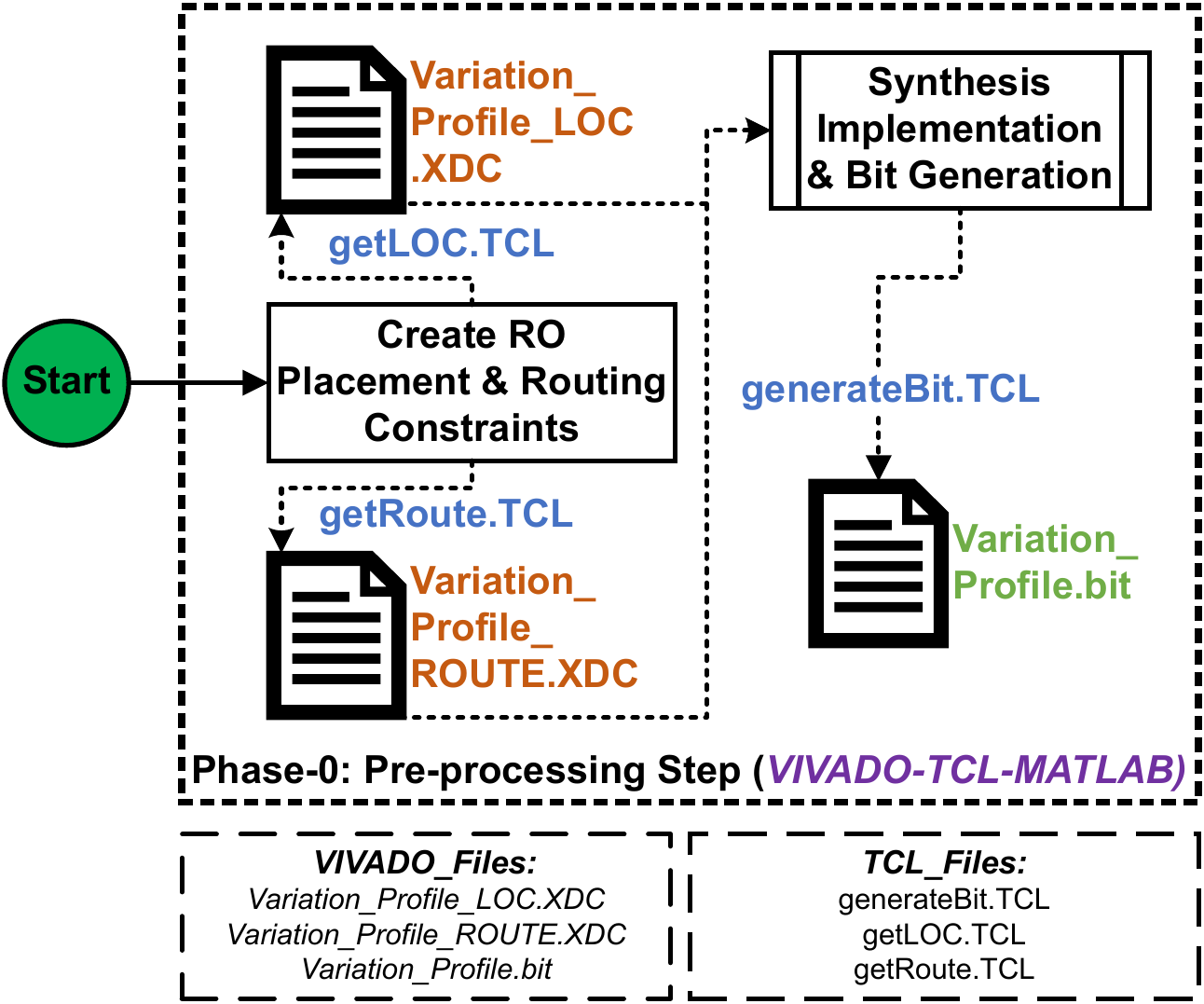}\caption{ \textit{MATLAB}, \textit{TCL} and \textit{VIVADO} interface has been used for device constraint creation and bit-file generation for characterization phase. There are two \textit{VIVADO} constraint files, and one bitstream file has been generated using three \textit{TCL} scripts.
\label{fig:phase_0}}
\end{figure}

The characterization phase requires to place $\mathcal{Z}$ number of ring oscillator to the entire FPGA chip, each ring oscillator requires a single slice for implementation (four slices for logic element and single shift register for latch). A software interface has been designed to automate the characterization constraint creation phase using \textit{MATLAB}, \textit{TCL} \& \textit{VIVADO} \cite{VIVADO_TCL_Interface}, as shown in Figure \ref{fig:phase_0}. The preprocessing phase requires to perform only once for the same type of FPGA chips; for example, all Nexys-4DDR FPGA requires only one constraint creation phase.

\subsubsection{Placement \& Routing}

The ring oscillators have been placed over the entire FPGA region, and \textit{XDCMacro} with detailed routing techniques are used to provide control over place \& route. Ring oscillators are constrained not to occupy the central region of FPGA fabric because we have appropriated it for other required circuitry. 

\begin{figure}[tbh]
\centering{}\includegraphics[width=0.8\columnwidth]{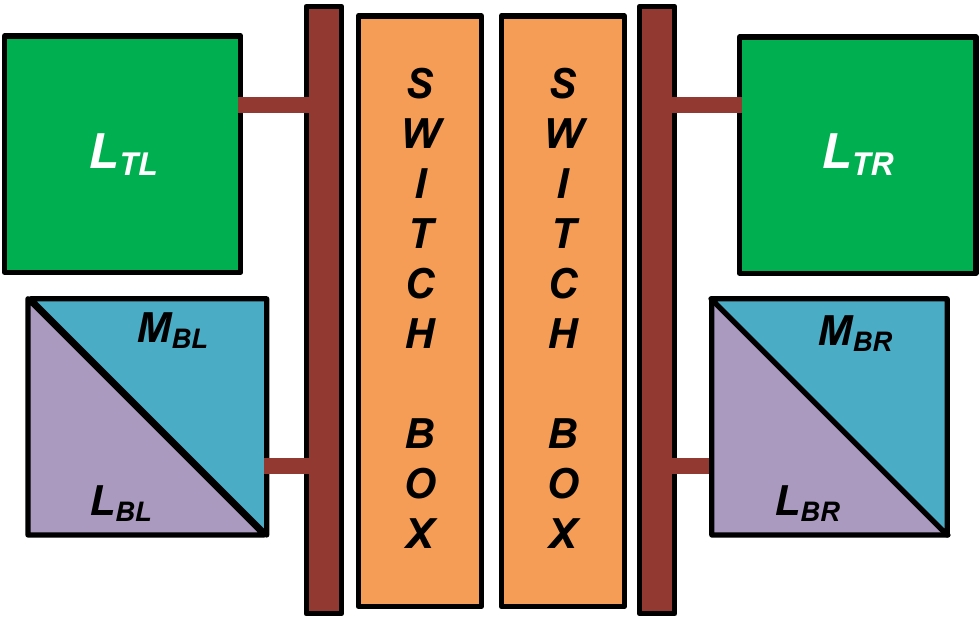}\caption{ Configurable logic block (CLB) architecture for \textit{Xilinx-7} series FPGA device with different configurations. Here, $L_{TL}, L_{TR}, L_{BL}, L_{BR}$ represents the slice-L at top-left, top-right, bottom-left, bottom-right. Similarly, $M_{BL}$ and $M_{BR}$ represents the slice-M at bottom-left and bottom right locations.
\label{fig:placement_config_JETTA}}
\end{figure}

The architecture of the Xilinx-7 series FPGA consists of configurable logic blocks (CLB), where each CLB consists of mainly two types of slices, i.e. L and M, and each one provides different routing paths \cite{VIVADO_CLB}. The placement location of each slice affects the routing path. There are mainly four possible configurations of the slice according to the physical location of the slice w.r.t. switch box. The slice connected to top side is always a \textit{L} and the bottom connected slice can be \textit{L} or \textit{M}. The slice connected to top-left, bottom-left, top-right and bottom-right is labeled as TL, BL, TR and BR, respectively, The different configurations of CLB are mentioned in Figure \ref{fig:placement_config_JETTA}.

The other slice classification is based on the routing delay of the ring oscillators. There are three possible sets related to the routing delay,  i.e. $L_{12}$, $L_{3}$ and $M$. The set $L_{12}$ contains slice-L which is placed at top side (i.e. $\text{L}_{TL}$ and $\text{L}_{TR}$). Similarly, set $L_{3}$ contains $\text{L}_{BL}$ and $\text{L}_{BR}$ locations, and set $M$ contains $\text{M}_{BL}$ and $\text{L}_{BR}$.

\begin{equation}
	\begin{split}
	L_{12} = L_{TR} \bigcup L_{TL}, \text{ } L_{3} = L_{BL} \bigcup L_{BR} \text{ and } \\ M = M_{BL} \bigcup M_{BR}
	\end{split}
\end{equation}

The routing path of each set should produce different routing delay.

\subsection{Phase-1: Frequency Variation Profiling}

The frequency variation profiling phase requires two steps; 1) Data extraction and 2) Post processing. The data extraction phase extracts the frequency, and the collected frequencies have been processed to remove the data transmission error.

\subsubsection{Data Extraction}

The first step for the proposed approach is data extraction by employing on-chip monitors, which can be performed with the help of hardware-software co-interface. The hardware part consists of an FPGA device, and the software interface has been designed \textit{MATLAB}. The data transmission is performed by employing standard \textit{UART} communication protocol with a data transmission speed of 115200 bits/seconds. The extracted count $(\alpha_{ro})$ data is converted to frequency data ($F_{ro}$) using (\ref{eq:delay_equation}). The mean value and standard deviation of the collected frequency data have been represented as $F_{ro_{\mu}}$ and $F_{ro_{\sigma}}$, respectively. All the collected data will be discarded after the completion of the second phase, to minimize the risk of information leakage.

\subsubsection{Erroneous RO Rejection}
{\color{blue}
The second step performs the post-processing that has been used to remove the highly deviated ring oscillator frequencies. These highly deviated frequencies are termed as "erroneous RO frequencies". Moreover, we have rejected these frequencies and the obtained error-free ring oscillator frequencies ($F$) using (\ref{eq:Unstable_ROs}).

\begin{equation}
F = F_{ro}\left( \frac{ f_{ro_{\sigma}}^x }{ f_{ro_{\mu}}^x } \leq Th \right), \forall  \hspace{1pt}  x  \hspace{1pt}  \epsilon  \hspace{1pt}  F_{ro}\label{eq:Unstable_ROs}
\end{equation}

Here, $F_{ro}$ is the ring oscillator frequencies obtained from the characterization phase and $(F)$ represents the error-free ring oscillator frequencies. The mean and standard deviation value of $x^{th}$ RO frequency is represented as $f_{ro_{\mu}}^x$ and $f_{ro_{\sigma}}^x$, respectively.}

The parameter $Th$ is defined as the maximum allowable value of the normalized standard deviation. We have used normalized standard deviation with mean value to reduce the effect of mean value on the standard deviation. We have analyzed the data of $54$ FPGA chips and observed that the maximum of $1\%$ ring oscillator frequencies are showing bit flipping error and some frequencies are showing high standard deviation, which can produce unreliable response bits. Therefore, we have fixed $Th$ such that $5\%$ of the total frequencies are discarded and not used for further processing.

\begin{figure}[tbh]
\centering{}\includegraphics[width=1\columnwidth]{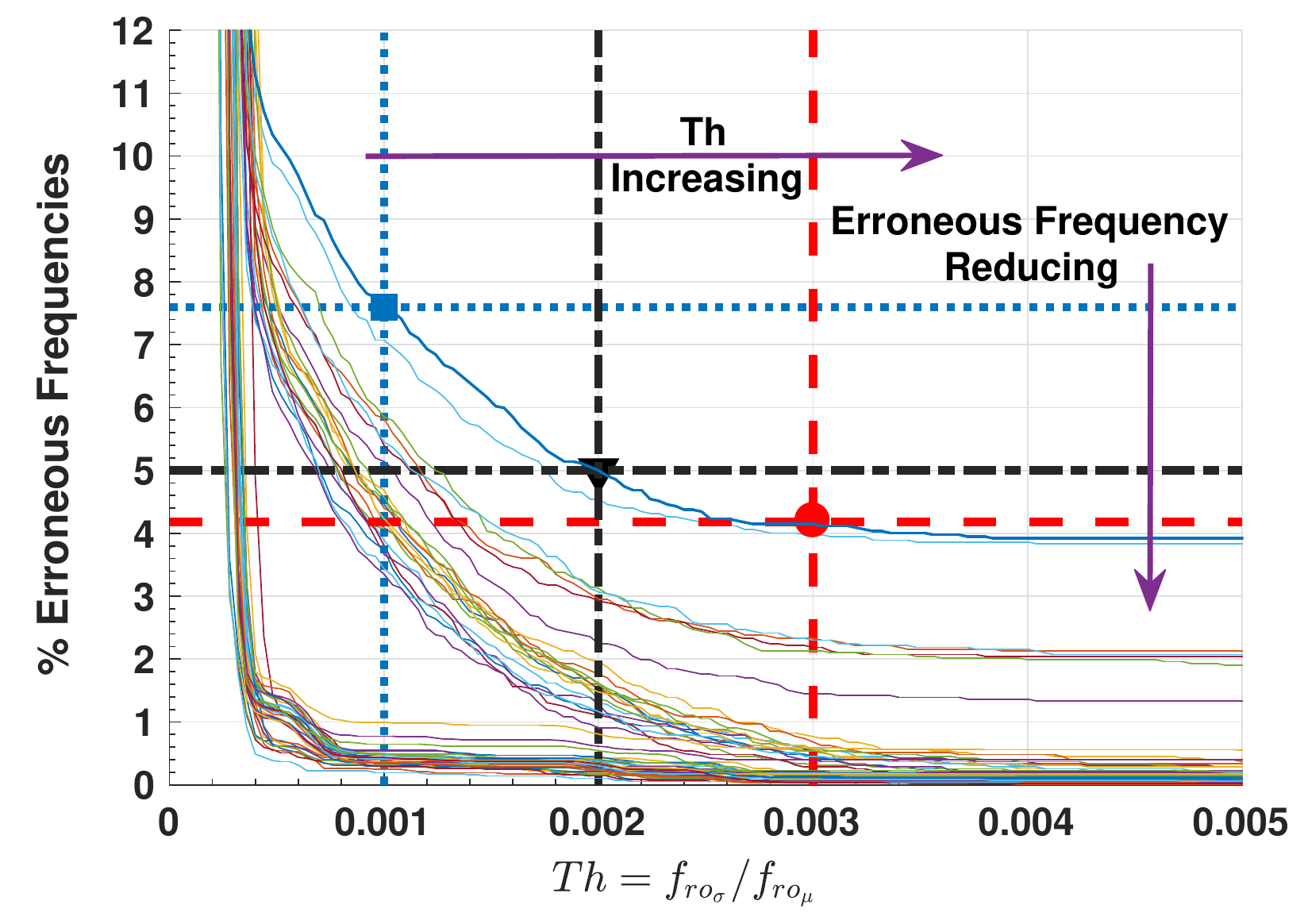}\caption{The effect of threshold selection on the number of erroneous ring oscillator for 54 FPGA chips. The rejection threshold is fixed at $Th = 0.002$, which can reject a maximum of 5\% ring oscillator instances. 
\label{fig:erroneous_ro_rejection}}
\end{figure}

{\color{blue}

The effect of threshold variation ($Th$) on the number of the erroneous ring oscillator is depicted in Figure \ref{fig:erroneous_ro_rejection}. We have achieved a threshold value $Th = 0.002$ for $5\%$ error band. However, the increment in $Th$ value reduces the \% erroneous frequencies and vice versa. 
}

\begin{figure*}
\centering
\subfloat[][\textbf{Nexys-4 DDR (Artix-7)}]{\includegraphics[trim = 1cm 0cm 0cm 0cm, width=0.65 \columnwidth]{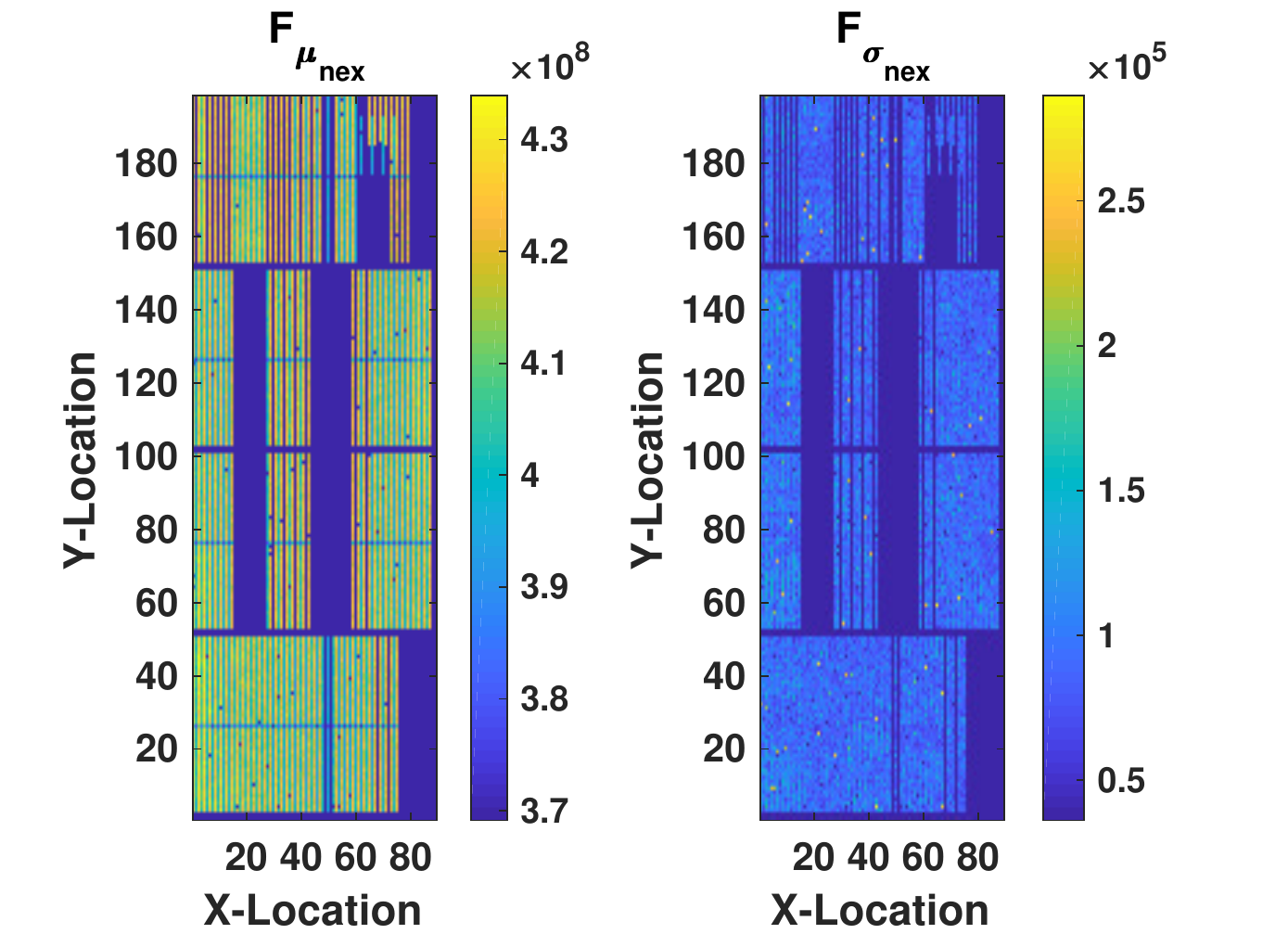} \label{fig:VMAP_NEX} }
\subfloat[][\textbf{Basys-3 (Artix-7)}]{\includegraphics[trim = 1cm 0cm 0cm 0cm, width=0.65 \columnwidth]{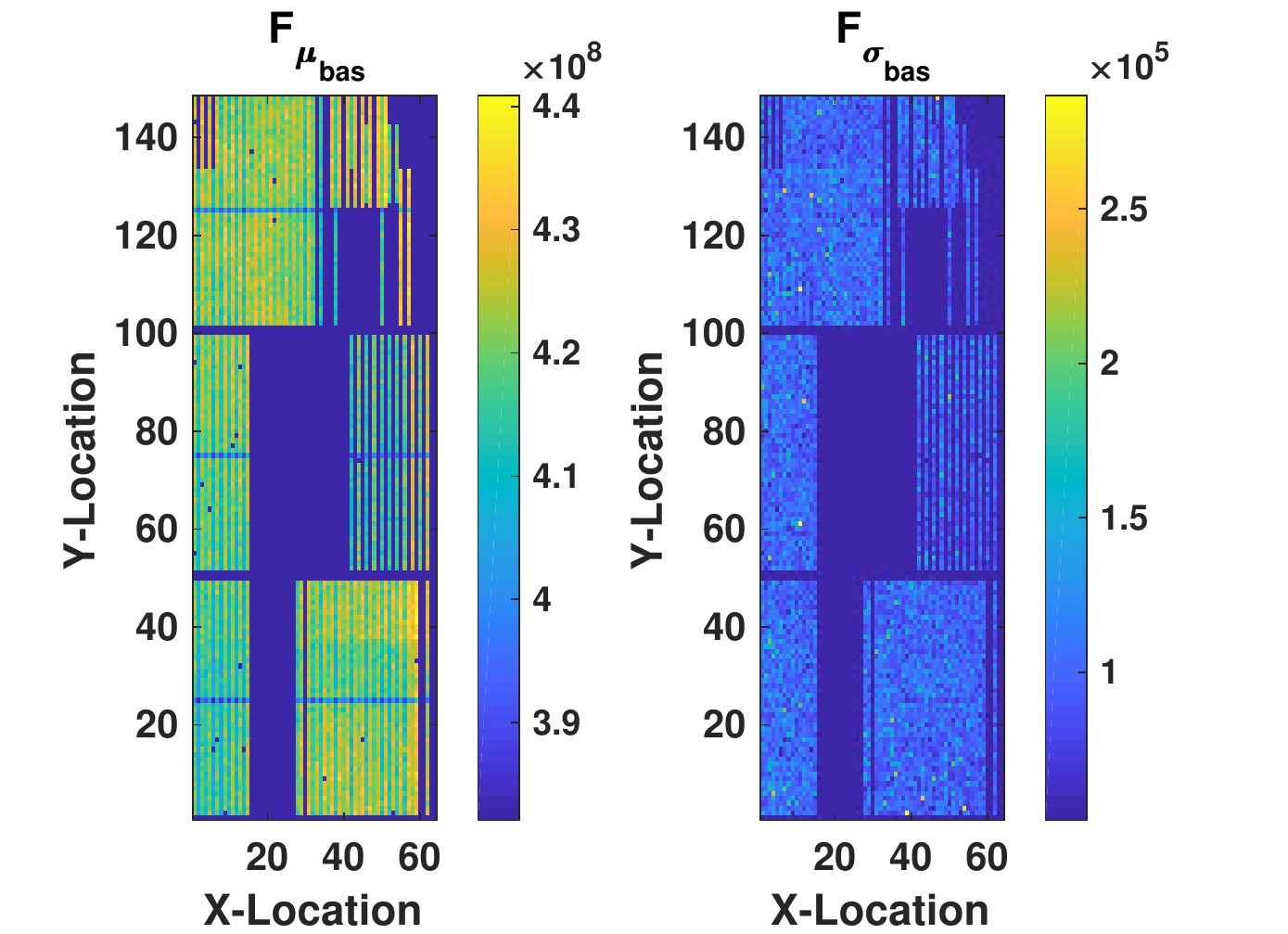}\label{fig:VMAP_BAS}}
\subfloat[][\textbf{Zybo (ZYNQ-7000)}]{\includegraphics[trim = 1cm 0cm 0cm 0cm,  width=0.65 \columnwidth]{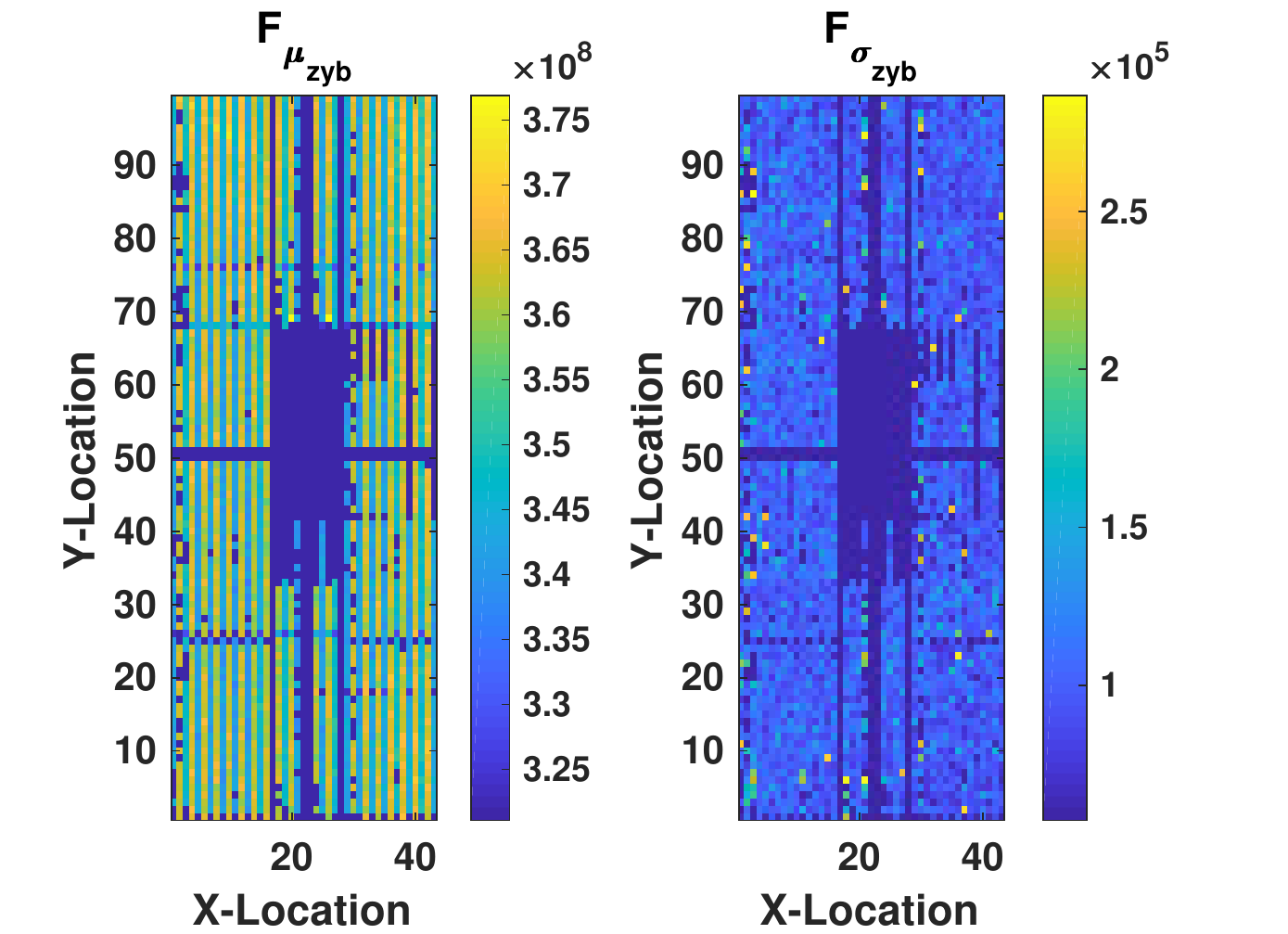}\label{fig:VMAP_ZYB}} 
\caption{\label{fig:Variation_Profiling_Data_JETTA} Frequency variation profiling data evaluated for three different FPGA chips. Here, $F_{\mu_{dtype}}$ and $F_{\sigma_{dtype}}$ are the mean and standard deviation value heat-map for $dtype$ FPGA device. Here, $dtype$ is $nex$, $bas$ and $zyb$ for \textit{Nexys-4DDR}, \textit{Basys-3} and \textit{Zybo}, respectively.}
\end{figure*}

% The reduction in $Th$ increases the \% of erroneous ring oscillators. $Th$ can be configured according to the ring oscillator frequency variation.

The selected error-free ring oscillator frequencies are represented as $F$, the mean values are $F_{\mu}$, the standard deviation values are $F_{\sigma}$ and the number of error-free frequencies are represented as $\overline{\mathcal{Z}}$

\subsubsection{Frequency Variation Results}

We have extracted a total of $11264$, $5696$ and $3520$ ring oscillators count data from each \textit{Nexys-4 DDR}, \textit{Basys-3} and \textit{Zybo} FPGA chips respectively at room temperature. We have collected a total of $32$ frequency samples for each ring oscillator. Figure \ref{fig:Variation_Profiling_Data_JETTA} shows the heat map for the mean ($F_{\mu}$) and standard deviation ($F_{\sigma}$) of frequencies extracted from all three type FPGA chips.

The mean frequency span $F_{\mu_{span}}$ for \textit{Nexys-4 DDR}, \textit{Basys-3} and \textit{Zybo} chips are  $64.78$MHz, $54.28$MHz and $54.71$MHz, respectively. Similarly, standard deviation span $F_{\sigma_{span}}$ are $249.4$KHz, $235.24$KHz and $229.4$KHz for \textit{Nexys-4 DDR}, \textit{Basys-3} and \textit{Zybo} FPGA chips, respectively.

\begin{figure}[tbh]
\centering{}\includegraphics[width=1\columnwidth]{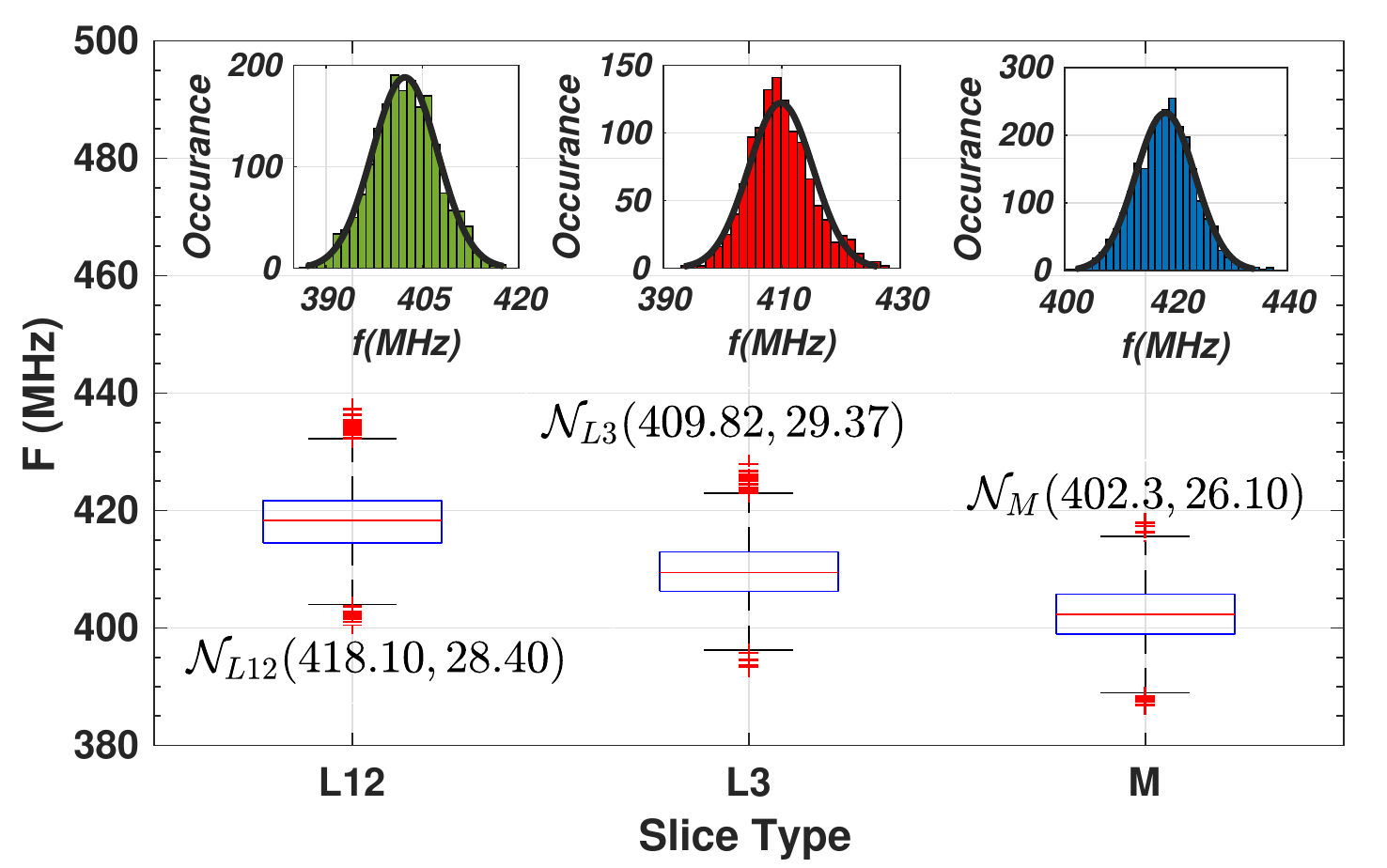}\caption{The measured frequency for different Slice locations after applying erroneous RO rejection on Basys-3 FPGA chip. The empirical mean and standard deviation for each type of slice have been evaluated using $\mathcal{N}(\mu, \sigma^2)$. 
\label{fig:routing_variation_VMAP}}
\end{figure}

\subsubsection{Effect of Routing Bias}
The reduction in routing delay increases the frequency and vice-versa. Figure \ref{fig:routing_variation_VMAP} shows the effect of routing delay variation on all three possible placement location sets for \textit{Basys-3} FPGA device. There are three different distributions, and each represents a slice type. The statistical distribution parameters have been evaluated with the help of empirical normal distribution of frequencies $\mathcal{N}(\mu, \sigma^2)$. The highest mean frequency $418.10$ MHz has been achieved by selecting locations from set $L_{12}$, and the minimum mean frequency $402.3$ MHz is achieved by selecting locations from slice-M. The maximum frequency range for biased placement and unbiased placement are $47.11$ and $32.51$ MHz, respectively. The improvement in the frequency range for biased placement is about $1.43$ times of the unbiased placement.

The ring oscillator implementation to all possible locations (i.e. $L_{12}, L_{3}$ and $M$) introduces routing bias, which increases frequency variation range. Therefore routing delay variation with biased placement certainly improved the frequency difference.

\subsection{Phase-2: ROPUF Creation}
The obtained set of error-free ring oscillator frequencies $(F)$ are randomly distributed across the frequency scale, and we have to select $M$ ring oscillators out of $\overline{\mathcal{Z}}$ number of ring oscillators to increase frequency difference. Therefore second phase adaptively searches for $M$ suitable locations out of $\overline{\mathcal{Z}}$, which maximizes the pairwise frequency difference (PFD). The frequency difference for a pair of ring oscillator is expressed as (\ref{eq:environmental_conditions_1}).

\begin{equation}
	\begin{split}
		\Delta f = \Delta \overline{f} + \Delta f_{\varepsilon_{m}} + \Delta f_{\varepsilon_{t}}
	\end{split}
	\label{eq:environmental_conditions_1}
\end{equation} 

Therefore, the essential condition required to avoid unstable response generation for each pair is manifested in (\ref{eq:necessary_condition_1}).

\begin{equation}
	\lvert \Delta \overline{f} \rvert \geq max\left(\lvert \Delta f_{\varepsilon_{m}} + \Delta f_{\varepsilon_{t}} \rvert \right)
	\label{eq:necessary_condition_1}
\end{equation}

The behavior of each ring oscillator is uncertain, therefore the estimation of $max\left(\lvert \Delta f_{\varepsilon_{m}} + \Delta f_{\varepsilon_{t}} \rvert \right)$ is difficult. Consequently, we have designed an adaptive approach to increase $\Delta \overline{f}$, such that it can compensate both environmental error as well as measurement error. Although measurement error is limited by employing erroneous RO rejection technique, still there are chances that aggregation of both errors can produce unreliable response bits.

\begin{figure}[tbh]
\centering{}\includegraphics[trim = 0cm 0cm 0cm 0cm, width=1\columnwidth]{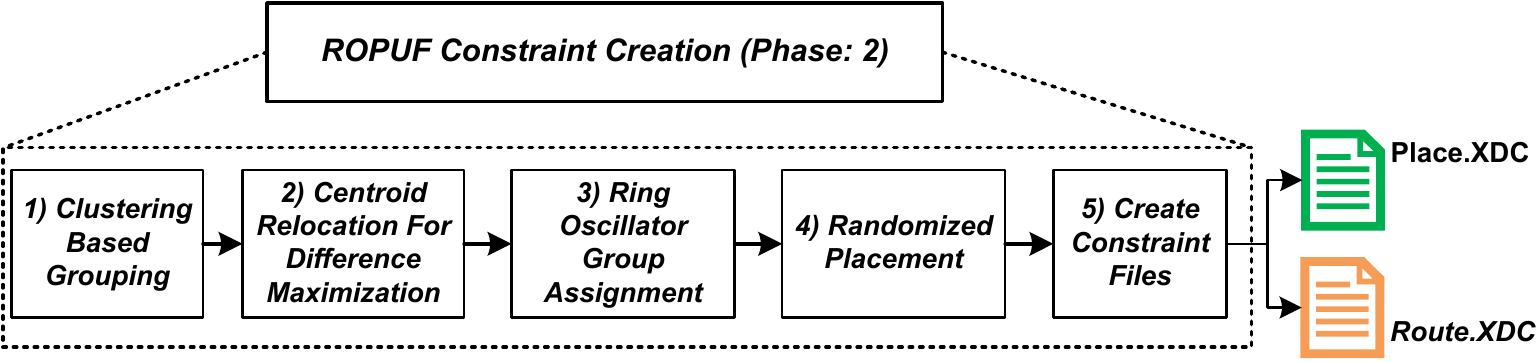}\caption{ ROPUF Constraint creation phase 2, this phase requires five steps to create place \& route constraints.  \label{fig:Proposed_Phase_2}}
\end{figure}

The approach, we are using is based on grouping such that it increases the inter-group frequency difference. The step used to create ROPUF place and route constraints are mentioned in Figure \ref{fig:Proposed_Phase_2}.

We are using K-means clustering based approach which has been used to create groups, such that the intra-group distance should be minimized \cite{Arthur07kmeans}. The reduction in intra-group distance may improve the inter-cluster distance. There are mainly two reasons to select K-means based clustering, which are as follows.

\begin{itemize}
	\item The number of frequencies to be selected ($M$) is fixed, therefore the number of clusters is the same as the number of frequencies to be selected.
	\item K-means clustering is sensitive to outliers (frequencies which are far apart from the mean value), and inclusion of noisy frequencies significantly improved the frequency difference.
\end{itemize}

We are assuming that K-means centroid frequencies ($\overline{C}$) during each iteration seems not to resemble at the frequency scale ($F$), therefore we are choosing the $i^{th}$ frequency which is near to the centroid ($\overline{c}_{i}$) using (\ref{eq:nearest_freq_selection}).

\begin{equation}
	\begin{split}
		\overline{c}_{i} = F (\text{indexof}(min{\lvert F - \overline{c}_{i} \rvert}) ), \forall i\in {1, 2, \dots, \overline{\mathcal{Z}}}
	\end{split}
	\label{eq:nearest_freq_selection}
\end{equation}

The inter-group frequency difference is further improved using centroid relocation based approach. The centroid relocation based approach shift the centroid and searches for the frequencies, which increases minimum pairwise frequency difference $(\chi)$. The minimum pairwise frequency difference ($\chi$) is defined as the minimum frequency difference among ${M\choose 2}$ possible pairs, and it can be evaluated using (\ref{eq:mpfd}).

\begin{equation}
	\chi = min(\lvert \Delta \overline{f}_{i} \rvert), \forall i \in \lbrace 1, 2, \dots {M\choose 2} \rbrace;
	\label{eq:mpfd}
\end{equation}

The metric $\chi$ provides a worst-case analysis of frequency difference for the proposed approach. Five steps have been used to generate ROPUF place \& route constraints. As we have modified the ring oscillator design by inserting a latch circuit, therefore we have omitted the routing path validation scheme from the previous work \cite{ARJUN_VLSID}. The following steps have been used during the second phase to select suitable ring oscillator locations, which improves $\chi$.

\subsubsection{Clustering Based Grouping}

K-means clustering approach works based on expected maximization approach, which is used to minimize the intra-cluster distance \cite{Dempster77maximumlikelihood}, and it may improve the inter-cluster distance. We are searching for the set of frequencies, which improves the minimum pairwise frequency difference ($\chi$). We have added a function in the K-means clustering to test the minimum pairwise frequency difference ($\chi$) during each iteration and stored the centroid values, which provides the highest $\chi$ among iterations. The convergence of the improved algorithm is the same as the standard K-means algorithm based on intra-cluster distance. The steps require to maximize $\chi$ has been explained in  Algorithm \ref{alg:grouping_approach}.

\begin{algorithm}[tbh]
\small
\SetAlgoLined
\SetKwFunction{findLocation}{findLocations}
\SetKw{initialValues}{generateInitialCentroid($F$)}
\SetKw{Phasezero}{Initialization:} 
\SetKw{Phasetwo}{K-means Clustering} 
\begin{raggedright}
\Phasezero\\
\Begin{
\small
$\overline{C}$ = \initialValues;
}
\Phasetwo\\
\Begin{
	\small	
	\textbf{1:} Calculate inter centroid minimum distance ($\chi$) in between all centroid $\overline{c}_{j}$,   $j \in 1, 2, \dots, M$ and store $\chi$ in variable $\beta_{p}$ and store the corresponding centroid value in a list $\ell_{p}$. \\
	\vspace{0.5em}
	\textbf{2:} Compute distance $\psi_{i}$ for $i^{th}$ frequency instance $f_{i}$,  $i \in 1, 2, \dots, \mathcal{Z}$ with all centroids $\overline{C}$.\\
	\vspace{0.5em}
	\textbf{3:} Label $i^{th}$ instance with $j$ such that $\psi_{i}$ is minimum. \\
	\vspace{0.5em}
	\textbf{4:} Replace centroid $\overline{c}_{j}, j\in 1,\dots ,M$ with average of the frequencies labeled with $j$. \\
	\vspace{0.5em}
	\textbf{5:} Find nearest existing frequency for each $\overline{c}_{j}$ from $F$ and store it in list $\ell_{c}$.\\
	\vspace{0.5em}
	\textbf{6:} Calculate $\chi$ in between all elements of list $\ell_{c}$ and store it in $\beta_{c}$. \\
	\vspace{0.5em}
	\textbf{7:} \If{ $\beta_{c} > \beta_{p}$}{ 
		(i) $\ell_{p}$ = $\ell_{c}$ and $\beta_{p}$ to $\beta_{c}$
	}
	\Else{
		(i) No modifications.
	}
	
 	\vspace{0.5em}
	\textbf{8:} Repeat Steps (2) to (8) until $\overline{c}_{j}$ does not becomes fixed or the  maximum number of iterations $k_{max}$.\\
 	\vspace{0.5em}
 	\textbf{9:} The optimized centroid list is $\ell_{p}$ and maximized $\chi$ is $\beta_{p}$.
}
\end{raggedright}
\caption{Modified K-means based grouping approach}
\label{alg:grouping_approach} 
\end{algorithm}

\paragraph{\textbf{Initialization}}
K-means approach sensitive to the initialization, thus we have applied different centroid initialization techniques, i.e. Linearly spaced frequency vector, Uniform density based initialization, K-means based initialization (K-means++) and Random initialization. 
\begin{itemize}
\item \textbf{Linearly spaced centroid:} The linearly spaced vector is generated in the ring oscillator frequency range, where two consecutive frequencies have the same frequency difference.
\item \textbf{Uniform density based centroid:} Two consecutive frequencies have the same number of instances.
\item \textbf{K-means based initial centroid:} K-means itself has been used for initial centroid generation.
\item \textbf{Random initial centroids:} Choose initial centroid uniformly at random in a provided frequency range.
\end{itemize}

\begin{figure}[tbh]
\centering{}\includegraphics[width=0.9\columnwidth]{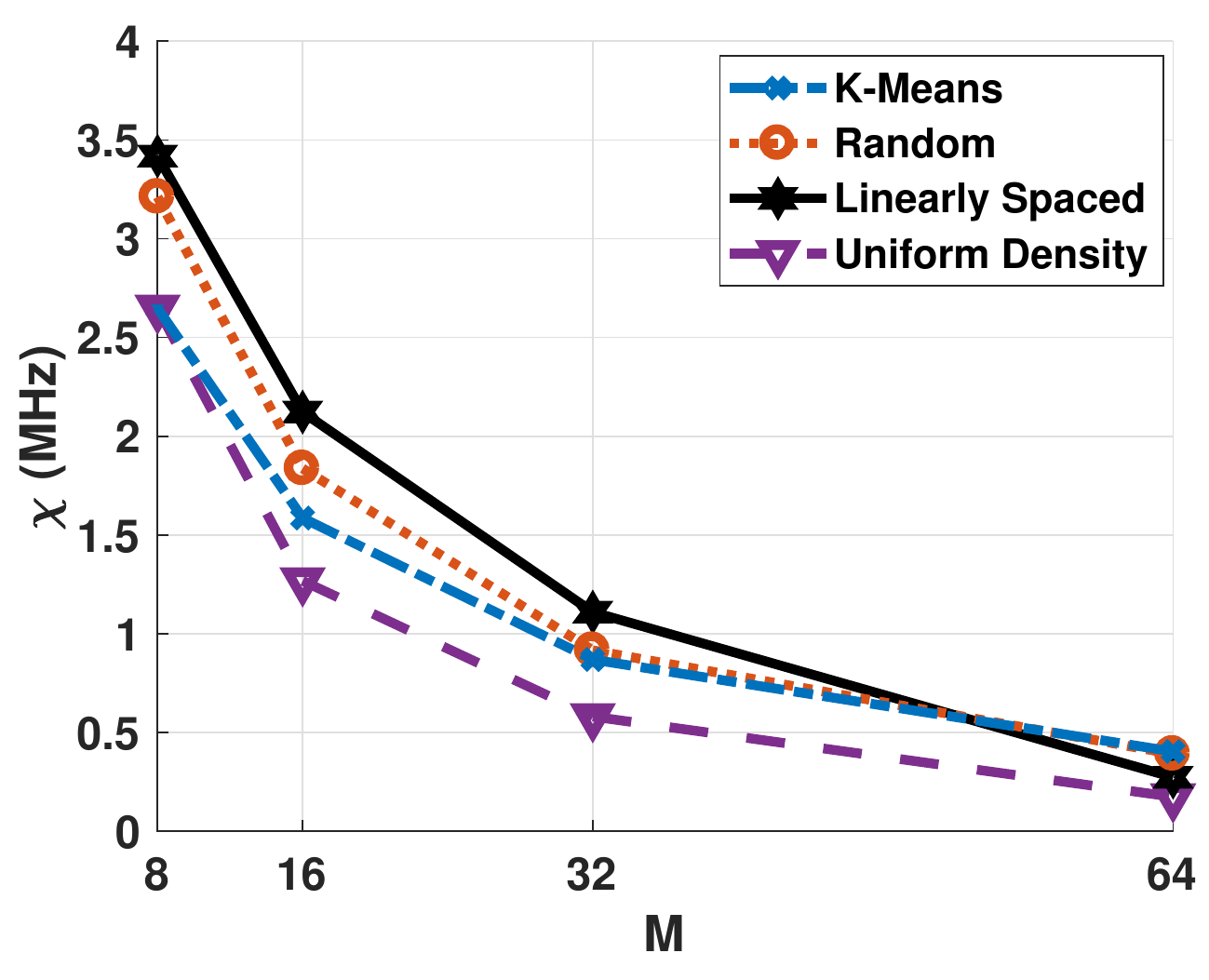}\caption{The variation of minimum pairwise frequency difference ($\chi$) for different seed initialization methods and ring oscillator instances ($M$).
\label{fig:K_Means_Based_seeding}}
\end{figure}

The effect of seeding on $\chi$ is shown in Figure \ref{fig:K_Means_Based_seeding}. The linearly spaced centroids based initial vector improves the $\chi$ for $M\leq 32$, but the performance degrades for $M>32$. Similarly, the performance for K-means based initialization provide better $\chi$ for $M\geq64$.

\paragraph{\textbf{Intra-group Minimization}}
The intra group minimization has been performed with the help of distance function $(\psi)$. The minimum distance $\psi_{i_{min}}$ for $i^{th}$ frequency instance with all centroids $\overline{C}$ has been evaluated using (\ref{eq:distance_formula}).

\begin{equation}
	\psi_{i_{min}} =  min \left( \lVert c_{j} - f_{i} \rVert  \right), \forall \overline{c}_{j} \in \overline{C}
	\label{eq:distance_formula}
\end{equation}

K-means based approach minimizes $\psi$ during each iteration and for each frequency instance. The assessment of K-means clustering has been performed by employing mean intra-cluster distance (MICD) \cite{ICMD_MICD}. The mean intra-cluster distance (MICD) for $j^{th}$ centroid can be determined using (\ref{eq:MICD_Formula}).

\begin{equation}
\begin{split}
\text{MICD}_j = \frac{1}{N_{j}} \sum_{x\in F_{\overline{c}_{j}}}  \lVert x-\overline{c}_{j} \rVert \\ \overline{MICD} = \frac{1}{M} \sum_{j=1}^M (\text{MICD}_j) 
\end{split}
\label{eq:MICD_Formula}
\end{equation}

Here $F_{\overline{c}_{j}}$ is the set of ring oscillator frequencies relating $\overline{c}_{j}$ centroid frequency, $N_{j}$ are the number of elements in $j^{th}$ cluster and $\overline{MICD}$ is the mean of MICD.

\paragraph{\textbf{Inter-group Maximization}}
The inter-cluster minimum distance (ICMD) has been used to evaluate the performance of K-means clustering \cite{ICMD_MICD}. The metric ICMD is same as minimum pairwise frequency difference ($\chi$) for K-means clustering. The evaluation of $\chi$ has performed during each iteration, and $\chi$ value along with the centroids has been stored in $\beta$ and list $\ell$, respectively. 

K-means convergence is the same as earlier to minimize intra-cluster distance, but we have modified the algorithm in such a way that it also checks the centroids, which can improve the $\chi$. The algorithm converges with the minimization of the mean intra-cluster distance. 

\begin{figure}[tbh]
\centering{}\includegraphics[width=1\columnwidth]{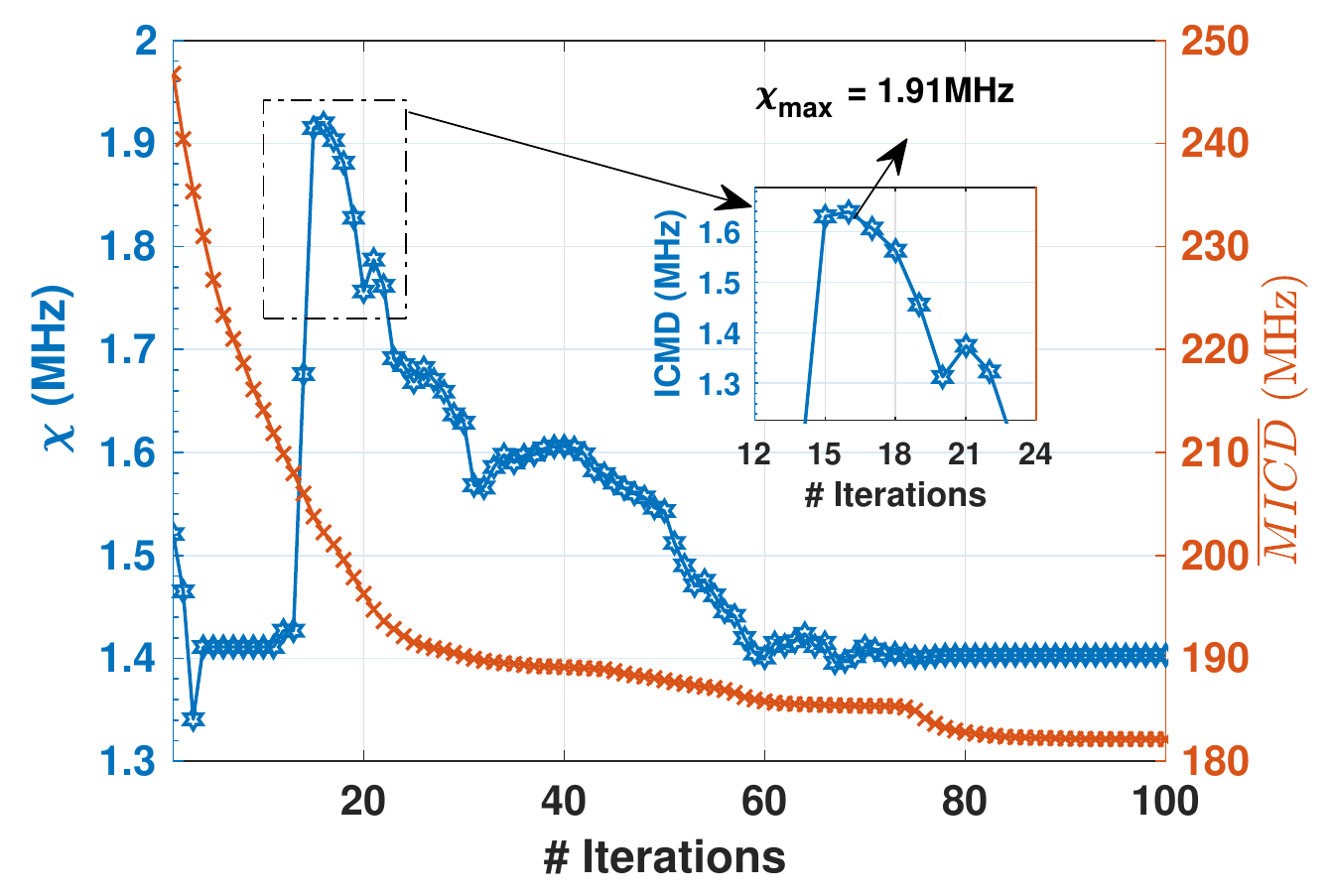}\caption{ 
The effect of improved K-means on $\chi$ during iterations. The maximum $\chi_{max}$ is achieved at iteration \#16, therefore optimized centroids are centroids available at iteration \#16.
\label{fig:ICMD_MICD_Iterations}}
\end{figure}

The proposed approach dependent on number of ring oscillator instances (M), therefore we have assessed the performance of K-means by employing $\overline{MICD}$ and $\chi$ evaluation metric. The effect of improved K-means on $\chi$ and $\overline{MICD}$ is depicted in Figure \ref{fig:ICMD_MICD_Iterations}. $\overline{MICD}$ metric improvement is similarly as standard K-means, but we are searching for the global maximum of $\chi$, which is achieved at iteration \#16. We are storing the centroid values belongs to the iteration \#16 and consider these centroids as optimized centroids.

The effectiveness of the improved K-means based grouping approach has been evaluated by comparing it with some other frequency selection techniques, i.e. Mean based selection, Median based selection, Random frequency selection and K-means based selection \cite{ARJUN_VLSID}.

\begin{itemize}
\item \textbf{Mean based selection:} Select the $M$ frequency data points such that the neighbor frequencies have an equal distance.
\item \textbf{Median based selection:} Select $M$ frequency data points such that the difference between two consecutive frequencies has the same number of ring oscillator instances.
\item \textbf{Random frequency selection:} Select $M$ frequency instances, uniformly at random from all the available frequency instances.

\end{itemize}

\begin{figure}[tbh]
\centering{}\includegraphics[width=1\columnwidth]{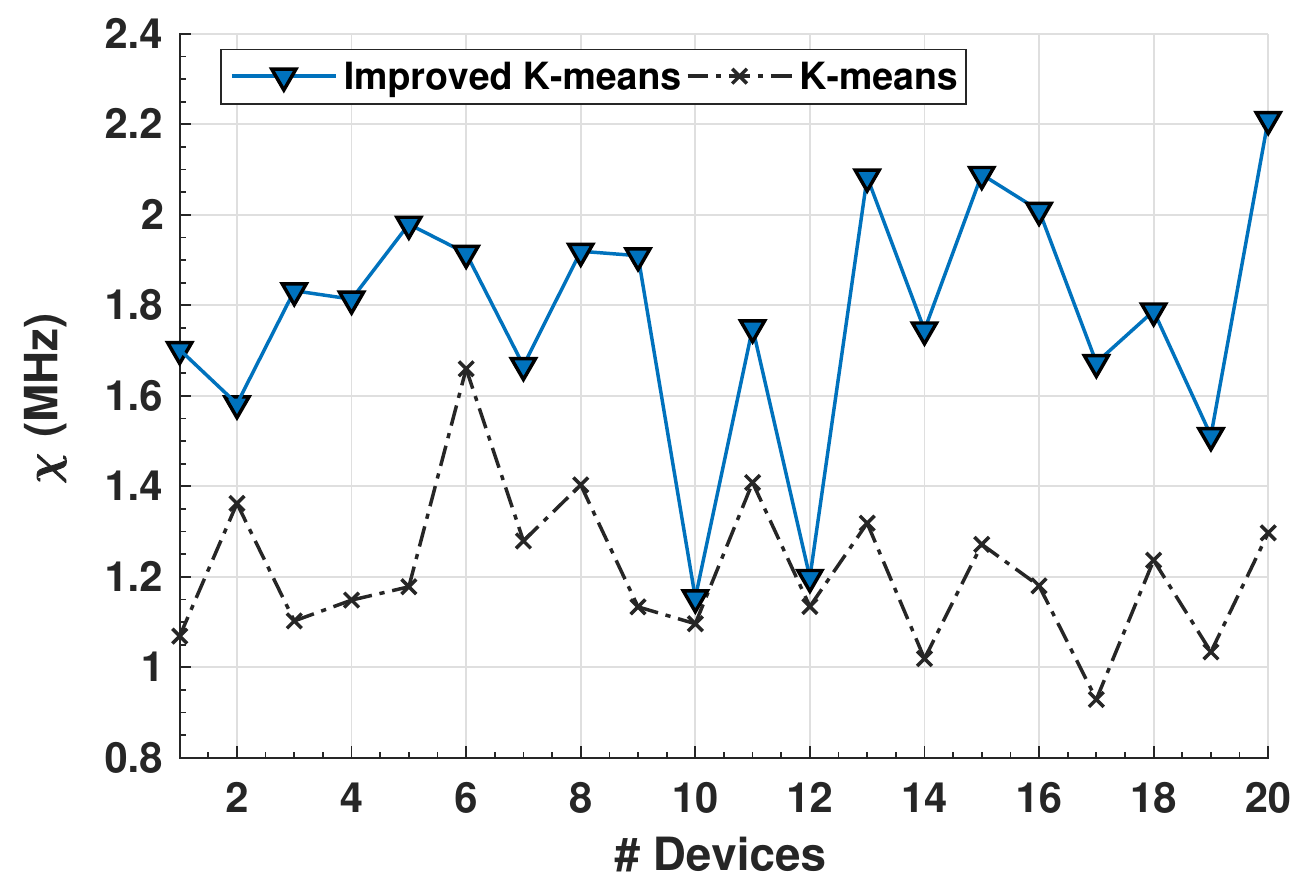}\caption{ 
The performance evaluation of improved K-means with the K-means in terms of minimum pairwise frequency difference ($\chi$) on $20$ \textit{Nexys-4DDR} FPGA devices and the ROPUF configuration is $M16$.
\label{fig:Kmeans_comparison_with_others}}
\end{figure}

We have improved K-means clustering by including inter-centroid global maximum selection technique along with mean based initialization. The frequency difference increment is evaluated by comparing the improved K-means approach with the other approaches. We have performed a comparison of improved K-means with previous K-means for $20$ \textit{Nexys-4 DDR} FPGA devices, as shown in Figure \ref{fig:Kmeans_comparison_with_others}. We have found that improved K-means provides an average improvement of $46.4\%$ as compared to K-means \cite{ARJUN_VLSID}.

\begin{figure}[tbh]
\centering{}\includegraphics[width=1\columnwidth]{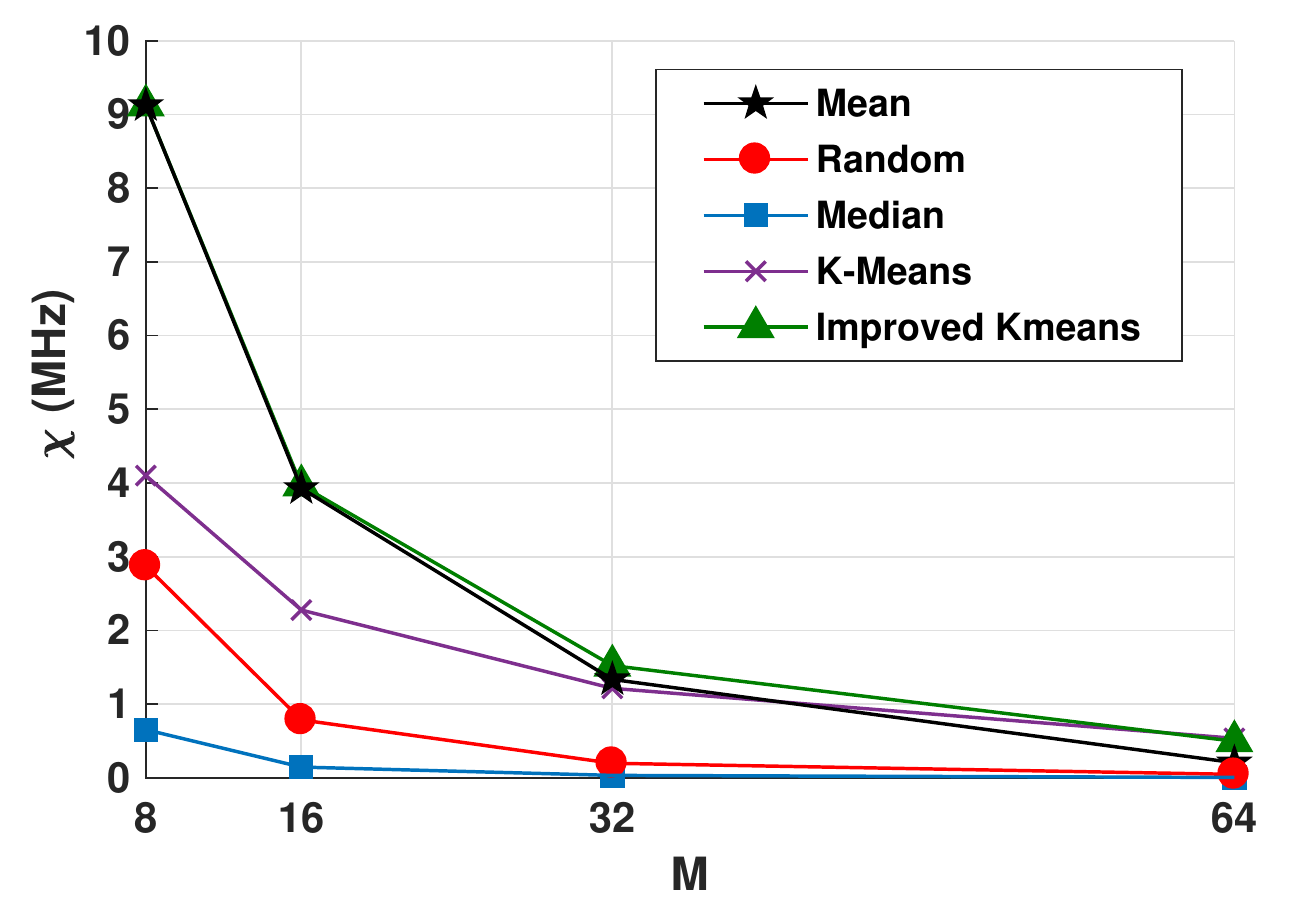}\caption{The comparative performance evaluation of the improved K-means with other frequency selection approaches in terms of $\chi$ and the variation of $\chi$ with M.
\label{fig:K_Means_Based_comparison}}
\end{figure}

The comparison with other approaches with improved K-means are shown in Figure \ref{fig:K_Means_Based_comparison}, which signifies that the improved K-means based approach increase the $\chi$ among all methods. The second best performance is the mean based frequency selection procedure. The proposed approach provide same frequency difference for $M\leq16$ and it improves the $\chi$ for $M>16$.

\begin{algorithm}[tbh]
\small
\SetAlgoLined
\SetKwFunction{findLocation}{findLocations}
\SetKwInOut{Input}{Input}
\SetKw{algostart}{Pseudocode:}
\SetKw{initialValues}{generateInitialCentroid($F_{RO}$)}
\SetKw{Phasezero}{Initialization:} 
\SetKw{Phasetwo}{K-means Clustering} 
\begin{raggedright}

\Input{Sorted ring oscillator frequency mean values = $\nu$ \\
Number of centroids = $M$\\
Maximum Iterations = $max_{iters}$	\\
K-means centroid list = $\ell_{p}$ \\
} 
\algostart \\
\Begin{
	\small	
	\textbf{1:} Perform sorting operation in ascending order on K-means centroid list $\ell_{p}$ and stored it in $\ell_{s}$ . \\
	\vspace{0.5em}
	\textbf{2:} Evaluate successive pairwise centroid frequency difference $\Delta\ell_{s_{k}}$, $k=(i,j), i \in 1, 2, \dots 2M-1$ and $j=i+1$. \\
	\vspace{0.5em}
	\textbf{3:} Find the minimum value $\Delta\ell_{s_{min}}$ and its index $k_{m}$ from frequency difference list $\Delta\ell_{s_{k}}$. Assign threshold $th$ to $\Delta\ell_{s_{min}}$.\\
	\vspace{0.5em}

	\textbf{4:} Select the neighbor frequency difference $\Delta\ell_{s_{k_{m}-1}}$ and $\Delta\ell_{s_{k_{m}+1}}$.\\
	\vspace{0.5em}

	\textbf{5:} Check neighbor frequency difference, \textbf{if} $\Delta\ell_{s_{k_{m}-1}} > \Delta\ell_{s_{k_{m}+1}}$ then centroid movement direction $dir = L$ \textbf{else} $dir = R$. $L$ = Left and $R$ = Right. \\
	\vspace{0.5em}
	
	\textbf{6:} if $dir = L$ then shifting element is $\ell_{s_{i}}$ from sorted frequency data $\nu$, else shifting elements is $\ell_{s_{j}}$.
	\vspace{0.5em}

	\textbf{7:} \If {$dir == L$}{\textbf{(i)} Find the index of $\ell_{s_{i}}$ and $\ell_{s_{i-1}}$ from frequency data $\nu$, and store it in $\hbar_{R}$ and $\hbar_{L}$, respectively. \\
	\vspace{0.5em}
	\textbf{(ii)} Search new $\overline{\ell}_{s_{i}}$ from $\nu_{\hbar_{R}}$ to  $\nu_{\hbar_{L}}$ such that the index increment factor $p$ should satisfy $\lvert \nu_{\hbar_{L}} - \nu_{\hbar_{R-p}} \rvert>th$, $p \in 1, 2, \dots, p_{max}$. \\
	\vspace{0.5em}
	\textbf{(iii)} If $p$ does not satisfy (ii), then reverse the direction for search $dir = R$ and goto step (6). \\  % Avoid to stuck at local maximum
	\vspace{0.5em}
	\textbf{(iv)} New centroid value $\overline{\ell}_{s_{i}}$ is $\nu_{\hbar_{R-p_{max}}}$.

	}
	\Else{
	\textbf{(i)} Find the index of $\ell_{s_{j}}$ and $\ell_{s_{j+1}}$ from frequency data $\nu$, and store it in $\hbar_{L}$ and $\hbar_{R}$, respectively. \\
	\vspace{0.5em}
	\textbf{(ii)} Search new $\overline{\ell}_{s_{j}}$ from $\nu_{\hbar_{L}}$ to  $\nu_{\hbar_{R}}$ such that the index increment factor $p$ should satisfy $\lvert \nu_{\hbar_{R}} - \nu_{\hbar_{L+p}} \rvert>th$, $p \in 1, 2, \dots, p_{max}$. \\
	\vspace{0.5em}
	\textbf{(iii)} If $p$ does not satisfy (ii), then reverse the direction for search $dir = L$ and goto step (6). \\  % Avoid to stuck at local maximum
	\vspace{0.5em}
	\textbf{(iv)} New centroid value $\overline{\ell}_{s_{j}}$ is $\nu_{\hbar_{L+p_{max}}}$.
	}
	\textbf{8:} Update current centroid $\ell_{s_{i}}$ or $\ell_{s_{j}}$ with $\overline{\ell}_{s_{i}}$ or $\overline{\ell}_{s_{j}}$, respectively. \\
	\vspace{0.5em}
	
	\textbf{9:} Repeat step (2) - (9) until location varies or till maximum number of iterations $max_{iter}$. \\
	\vspace{0.5em}

}
\end{raggedright}
\caption{Centroid relocation for difference maximization}
\label{alg:centroid_relocation} 
\end{algorithm}

\subsubsection{Relocating Centroid For Difference Maximization}
The centroid obtained through the modified K-means algorithm increases frequency difference but the difference is randomly distributed, and minimum pairwise frequency difference can be increased by shifting selected frequency points, therefore we are proposing centroid relocation based difference maximization approach as shown in Algorithm \ref{alg:centroid_relocation}.

The approach has been used after getting centroids from a modified K-means clustering approach. The approach evaluates frequency difference and checks the possible movement direction, If the frequency difference in the left side is larger than right side then shifting has to be performed in the left direction, else the direction will be the right side. The shifting can reduce the high-frequency difference and increases the low-frequency difference. The approach is iterative and performed until the possible movement locations does not exhaust.

\begin{figure}[tbh]
\centering{}\includegraphics[width=0.9\columnwidth]{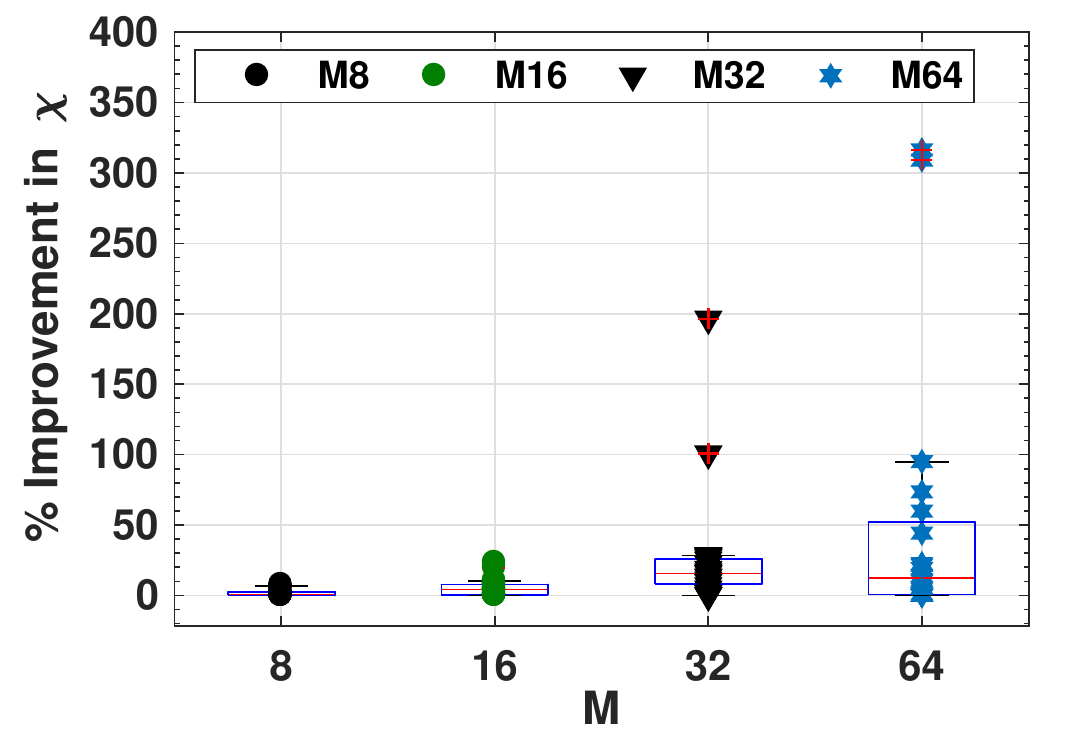}\caption{
The \% improvement in the $\chi$ by employing centroid relocation based approach and the effect of $M$ value on $\chi$.
\label{fig:K_Means_Based_Approach_Review}}
\end{figure}

The improvement in $\chi$ is evaluated for $20$ Nexys-4 DDR FPGA devices for $M=8, 16, 32 $ and $64$ ring oscillators. Figure \ref{fig:K_Means_Based_Approach_Review} shows the improvement in $\chi$ and the improvement is significant. Moreover, the average improvement for $M=8$ and $M=16$ are less than $10\%$, but the improvement for $M=32$ and $M=64$ are $27.69\%$ and $50.37\%$. The improvement is higher for a large value of M because, for large ring oscillators the K-means based approach provides small $\chi$ and centroid shifting improvement is comparatively significant.

\subsubsection{Ring Oscillator Group Assignment}
The hardware design comprises of two different groups; lower groups and upper groups, where all ring oscillator instances of the lower group (LG) should be compared to the ring oscillators at the upper group (UG). Each group contains $\frac{M}{2}$ ring oscillator frequencies, such that each group contains some frequencies in ordered and some frequencies in random trend. The ratio of $\kappa$ decides that the number of ring oscillator frequencies are randomly allocated with total frequencies. The possible value of parameter $\kappa$ can be evaluated using (\ref{eq:param_kappa}).

\begin{equation}
\begin{split}
	\kappa = \frac{i}{2^{x-1}} , \forall i\in \lbrace 0, 1, \dots, 2^{x-1} \rbrace, x = log_{2}\left( \frac{M}{2} \right)
\end{split}
\label{eq:param_kappa}
\end{equation}

Let's suppose that $M = 16$, then $x = 3$ and the possible randomness ratio $\kappa = \lbrace 0, 0.25, 0.5, 0.75, 1 \rbrace$. Therefore out of M ring oscillator frequencies, $\kappa\times M$ frequencies are randomly assigned and remaining $(1-\kappa)\times M$ are orderly assigned. The group assignment affects the uniformity, and two NIST statistical tests, i.e. frequency and block frequency test can provide a validation of uniformity, therefore we are validating ring oscillator assignment performance with NIST statistical tests.

\subsubsection{Randomized Placement}

The allocated ring oscillators frequencies follow the sorting order consequently we have performed group based randomization. We are randomly indexing selected ring oscillator frequencies. The ring oscillators are randomly placed, which deforms the output sequences in such a manner that the sequences should avoid creating the same response patterns on different hardwares. The randomized placement effectively removes the patterns, and it also reduces the chances for appearing the same sequence, therefore it enhanced the uniqueness as well as it also improves the randomness of the responses. The effect of randomized placement has been evaluated by employing the uniqueness metric and minimum entropy.

\subsubsection{Create Constraint Files }

The randomized placement provides the selected locations for each ring oscillators. We have extracted the routing path information from the characterization constraints and create placement and routing constraint files. The constraint creation is automated through \textit{MATLAB} and \textit{TCL} scripting.

\subsection{Algorithm Computation time}
The overall computation time is bifurcate in two parts; 1.) Frequency characterization phase ($t_{p1}$) and 2.) ROPUF constraints creation ($t_{p2}$). The synthesis implementation and bitstream generation time is not included for the fair assessment.
\subsubsection{Frequency Characterization Time ($t_{p1}$)}
The frequency characterization time ($t_{p1}$) is approximated linearly varied with the number of ring oscillator instances ($\mathcal{Z}$) and the number samples for each ring oscillators $(m)$ during characterization process. The ring oscillator enable pulse duration for each ring oscillator is fixed at $122.87 \mu sec$, and the data transmission time is considerable extensive as compared to enable pulse duration. The average time requires to measure a single sample of a ring oscillator instance is about $3msec$. We have unfolded circuit two times to reduce computation time. Moreover, it increases the characterization area, but the area will be unoccupied after the completion of the characterization phase.
\subsubsection{ROPUF Constraint Creation Time ($t_{p2}$)}
Apart from the characterization, the second phase also requires time to produce ROPUF place and route constraints. K-means clustering has a computation time complexity of $\mathcal{O}(n\log{}n)$ \cite{Arthur07kmeans}. Similarly, the computation time requires for the centroid relocation based approach has been evaluated using \textit{MATLAB} delay function. Figure \ref{fig:Proposed_Phase_2} shows the computation time dependency on the number of iterations for the centroid relocation approach, and it is evident that average computation time  increases with the increment in the number of average iterations. The computation time for constraint creation is much smaller than characterization $(t_{p2}\ll t_{p1})$, therefore it can be managed with characterization.

\begin{figure}[tbh]
\centering{}\includegraphics[width=1\columnwidth]{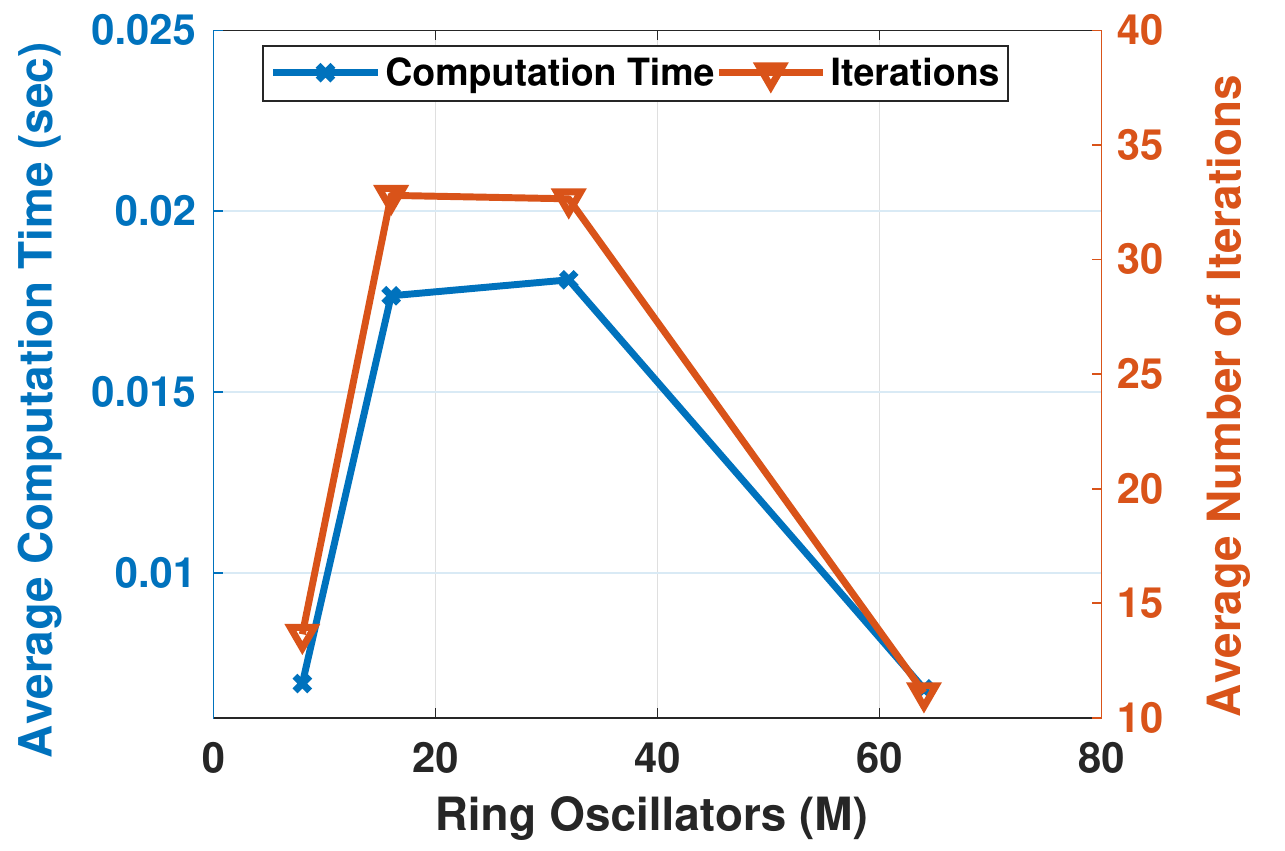}\caption{The effect of $M$ value on average computation time based on \textit{MATLAB} software and average number of iteration for centroid relocation approach.
\label{fig:Phase_2_computation_time}}
\end{figure}

\section{Experimental setup \& Evaluation Metric\label{sec:Experiment_results}}
The proposed method has been evaluated using standard quality metrics, i.e. uniqueness and reliability \cite{Quality_Metric_PUF}. The reliability and uniqueness metric have been evaluated by using intra-chip hamming distance $\text{HD}_{intra}$ and inter-chip hamming distance $\text{HD}_{inter}$. Hamming distance can be evaluated using (\ref{eq:HD_formula}).
\begin{equation}
	\begin{split}
		\text{HD}(x,y) = \sum_{z=1}^k \lvert R_{x_{z}} - R_{y_{z}} \rvert
	\end{split}
	\label{eq:HD_formula}
\end{equation}

Here, $R_{x}$ and $R_{y}$ are two $k$ bit response sequences. The randomness has been evaluated using NIST random statistical suite. Moreover, we have used a maximum of $54$ FPGA devices ( i.e. $24$ \textit{Nexys-4 DDR}, $10$ \textit{Basys-3} and $20$ \textit{Zybo}), and a single ROPUF has been implemented on a device, but the number of ROPUF per device can be increased by utilizing the remaining slices. We have used four different configurations, i.e. $M8$, $M16$, $M32$ and $M64$, whereas these configurations can produce $15$, $63$, $255$ and $1023$ bits long responses, respectively.

\subsection{Experimental Setup}

\begin{figure}[tbh]
\centering{}\includegraphics[width=0.7\columnwidth]{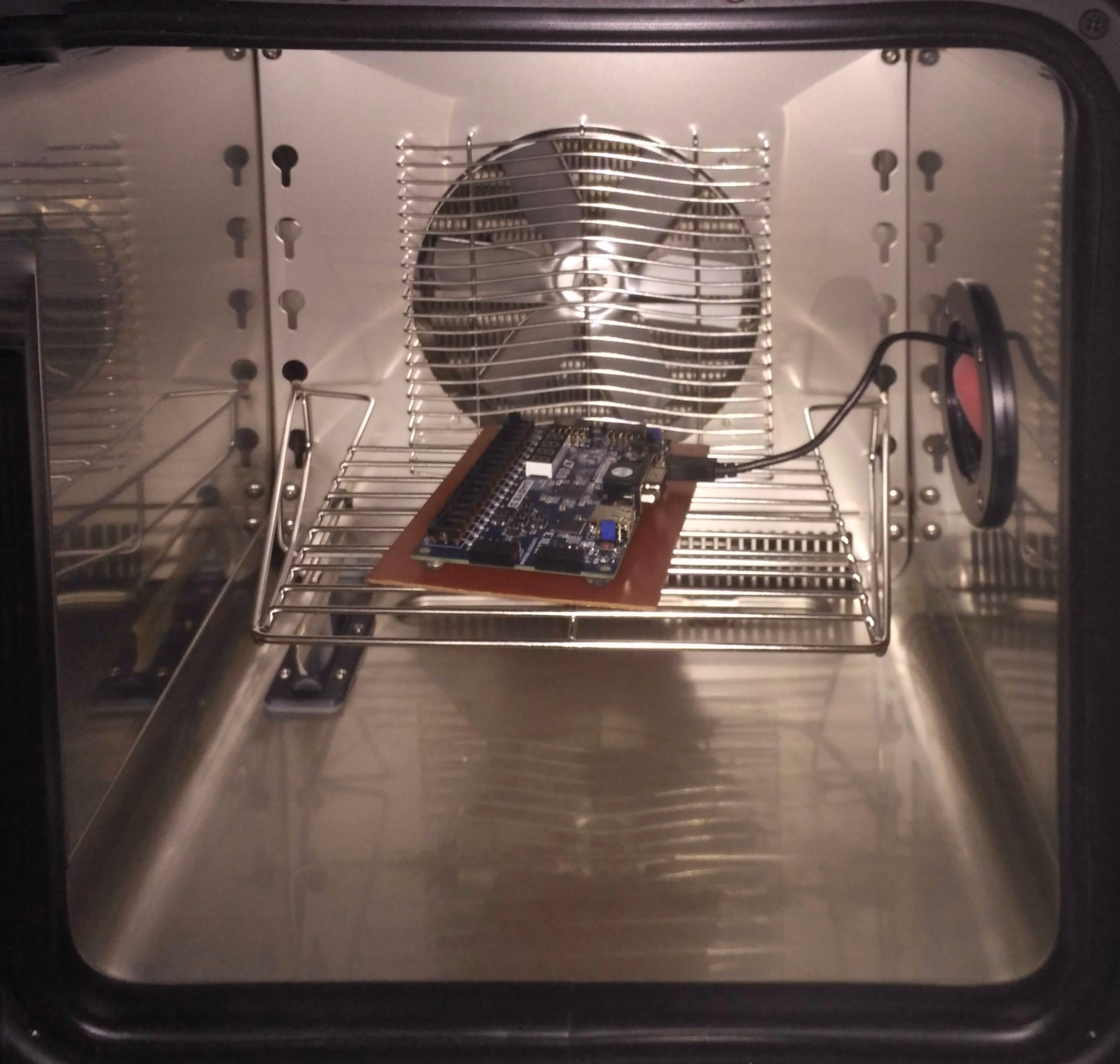}\caption{Temperature variation enclosure, \textit{"Espec SH-241 Temperature \& Humidity Chamber"} with a maximum variation range from $-40^\circ$ C to $150^\circ C$.
\label{fig:temp_setup}}
\end{figure}

We have designed "Hardware-Software" based co-interface for the automation of the characterization and ROPUF constraints creation phase by employing TCL and MATLAB scripts. The temperature variation is achieved with a temperature environment enclosure (\textit{Espec SH-241 Temperature \& Humidity Chamber}) as shown in Figure \ref{fig:temp_setup}, and the temperature range is fixed in between $-5^\circ C$ to $75^\circ C$ with a step size of $10^\circ C$. Similarly, supply voltage variation has been achieved by modifying one \textit{Basys-3} FPGA device, and the voltage variation range has kept in between $0.9V$ to $1.1V$ with a step size of $20$mV. The real-time temperature and the supply voltage has been monitored for verification, using \textit{XADC} interface and \textit{TCL} scripts. The reference operating conditions $(T_{ref}\text{=35}^\circ C$ and $V_{ref}\text{=1000}mV)$ has been used for the evaluation perspective.

\begin{figure}[tbh]
\centering{}\includegraphics[width=1\columnwidth]{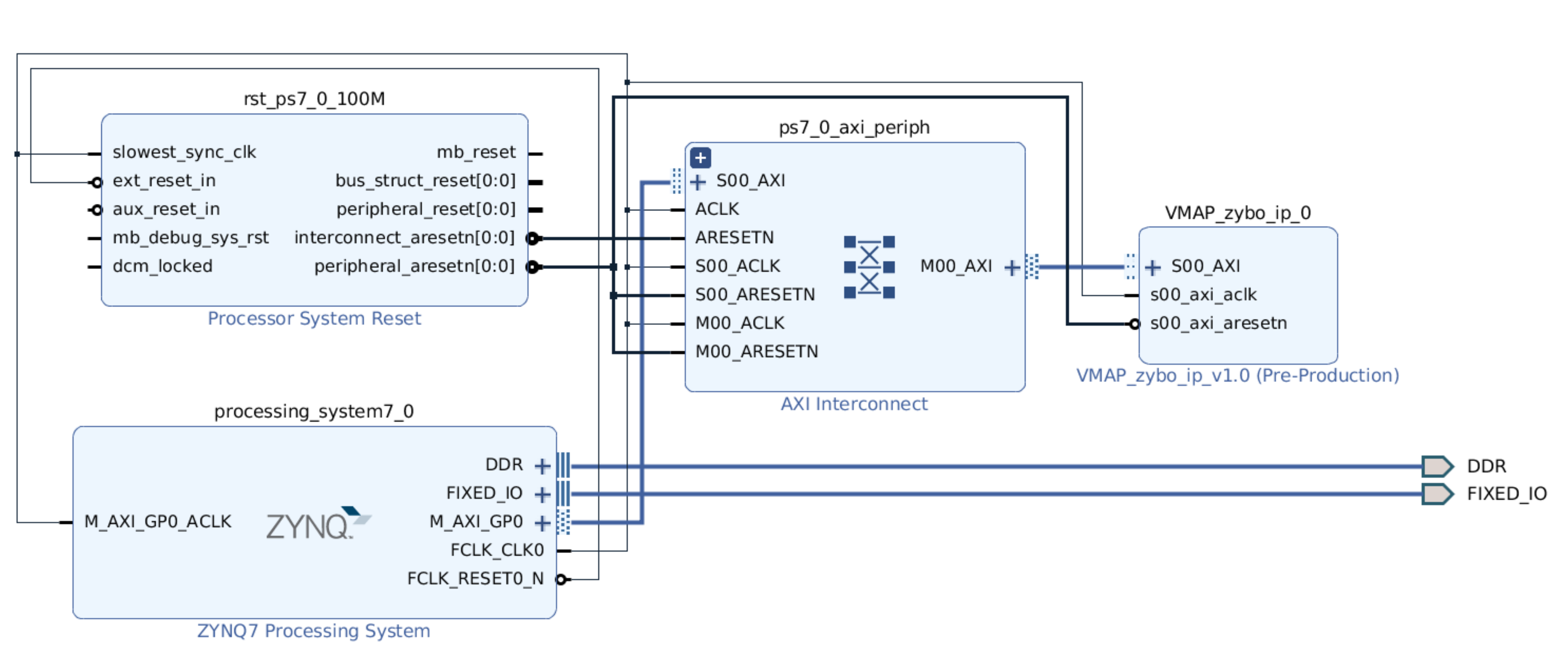}\caption{ Ring oscillator characterization and variation MAP creation using VMAP_zybo IP-Core, interfaced with AXI and \textit{ZYNQ-PS} (processing system).
\label{fig:zynq_PS_interface}}
\end{figure}

The proposed approach is completed in two different phases. The first phase is used to collect frequency data using on-chip frequency monitors. In order to extract frequency variations, we have designed and interfaced an IP-Core for characterization, termed as "VMAP_zybo" as shown in Figure \ref{fig:zynq_PS_interface}. The Zynq processing system IP-core has interfaced with AXI and UART-Lite for data extraction and data transmission \cite{ZYNQ_PS}. Similarly, we have implemented a standard UART protocol for \textit{Artix-7} FPGA devices.

% Please add the following required packages to your document preamble:
% \usepackage{multirow}
\begin{table*}[]
\caption{NIST statistical tests results variations with applicable $\kappa$ and $M$ values}
\renewcommand{\arraystretch}{1.5}
\small
\begin{tabular}{|l|c|c|c|c|c|c|c|c|}
\hline
\multicolumn{1}{|c|}{\multirow{2}{*}{\textbf{Test Name}}} & \multicolumn{2}{c|}{\textbf{M32 ($\kappa = 0.375$)}} & \multicolumn{2}{c|}{\textbf{M32 $(\kappa = 0.5)$}} & \multicolumn{2}{c|}{\textbf{M64 $(\kappa = 0.3125)$}}                           & \multicolumn{2}{c|}{\textbf{M64 $(\kappa = 0.375)$}}                            \\ \cline{2-9} 
\multicolumn{1}{|c|}{}                                    & \textbf{P-Value}          & \textbf{\% Proportion}       & \textbf{P-Value}         & \textbf{\% Proportion}      & \multicolumn{1}{c|}{\textbf{P-Value}} & \multicolumn{1}{c|}{\textbf{\% Proportion}} & \multicolumn{1}{c|}{\textbf{P-Value}} & \multicolumn{1}{c|}{\textbf{\% Proportion}} \\
\hline
\textbf{Frequency}                                        & 0.03517                   & 100.00 {\color{black}\cmark}                     & 0.07572                  & 98.14 {\color{black}\cmark}                    & 0.02054                               & 100.00 {\color{black}\cmark}                                   & 0.15376                               & 96.30 {\color{black}\cmark}                                    \\ \hline
\textbf{Block Frequency}                                  & 0.41902                   & 100.00 {\color{black}\cmark}                     & 0.65793                  & 100.00 {\color{black}\cmark}                    & 0.09657                               & 98.14 {\color{black}\cmark}                                    & 0.09731                               & 100.00 {\color{black}\cmark}                                   \\ \hline
\textbf{Cumsum Forward}                                   & 0.07572                   & 100.00 {\color{black}\cmark}                     & 0.02074                  & 98.14 {\color{black}\cmark}                    & 0.05898                               & 100.00 {\color{black}\cmark}                                   & 0.09657                               & 96.30 {\color{black}\cmark}                                     \\ \hline
\textbf{Cumsum Reverse}                                   & 0.01560                   & 100.00 {\color{black}\cmark}                    & 0.15376                  & 98.14 {\color{black}\cmark}                    & 0.28966                               & 100.00 {\color{black}\cmark}                                    & 0.19168                               & 96.30 {\color{black}\cmark}                                    \\ \hline
\textbf{Runs}                                             & 0.41902                   & 100.00 {\color{black}\cmark}                    & 0.12232                  & 100.00 {\color{black}\cmark}                   & 0.07572                               & 100.00 {\color{black}\cmark}                                   & 0.35048                               & 98.14 {\color{black}\cmark}                                    \\ \hline
\textbf{Longest Runs of 1's}                                     & 0.39706                   & 100.00 {\color{black}\cmark}                    & 0.81653                  & 100.00 {\color{black}\cmark}                   & 0.95583                               & 98.14 {\color{black}\cmark}                                    & 0.31908                               & 98.14 {\color{black}\cmark}                                    \\ \hline
\textbf{Entropy}                                          & 0.08558                   & 100.00 {\color{black}\cmark}                    & 0.57490                  & 100.00 {\color{black}\cmark}                   & 0.01304                               & 100.00 {\color{black}\cmark}                                   & 0.01451                               & 98.14 {\color{black}\cmark}                                    \\ \hline
\textbf{Serial}                                           & 0.53414                   & 100.00 {\color{black}\cmark}                    & 0.28966                  & 98.14 {\color{black}\cmark}                    & 0.17187                               & 100.00 {\color{black}\cmark}                                   & 0.13728                               & 96.30 {\color{black}\cmark}                                    \\ \hline
\textbf{Serial}                                           & 0.17186                   & 100.00 {\color{black}\cmark}                     & 0.17186                  & 100.00 {\color{black}\cmark}                   & 0.10879                               & 100.00 {\color{black}\cmark}                                    & 0.65793                               & 98.14  {\color{black}\cmark}                                   \\ \hline
\textbf{DFT}                                              & \textit{\textbf{NA}}      & \textit{\textbf{NA}}      & \textit{\textbf{NA}}     & \textit{\textbf{NA}}     & 0.02695                               & 100.00 {\color{black}\cmark}                                   & 0.45593                               & 100.00 {\color{black}\cmark}                                   \\ \hline
\end{tabular}

	\begin{tablenotes}
    \footnotesize
    	\item {\color{black}\cmark} : Passed Test
    	\item \textit{\textbf{NA}}: Not applicable
		\item Block frequency test block length (\textbf{M}): $20$ ,
		\item Max entropy block length \textbf{($m_e$)}: $m_e$ < $\lfloor log_{2}(n) \rfloor$ - 5
		\item Serial test block length \textbf{($m_s$)}: $m_s$ < $\lfloor log_{2}(n) \rfloor$ - 2
		\item Confidence interval $(\alpha_{test})$	: 0.01
	\end{tablenotes}

\label{tab:NIST_table}
\end{table*}

\vspace{-2em}
\subsection{Randomness Test}

{\color{blue}
The randomness of the response is evaluated by applying the NIST statistical test suite \cite{NIST_Test}. "A Statistical Test Suite for Random and Pseudorandom Number Generators for Cryptographic Applications" revision 1-A, has been used to test the randomness based on $\chi^2$ statistical distribution. The suite consists of a total of $15$ possible tests, and the applicability of each test is dependent on the input sequence length. Furthermore, each sequence should have at least $100$ bits to apply $9$ basic tests. The authors have validated the randomness effect on small length sequences \cite{NIST_Reviewer_1} \cite{Entropy_Distillers} \cite{FROPUF}. However, at least one million bit sequences required to apply all $15$ statistical tests. We have applied NIST tests to M16 and M32 configuration only, however, the remaining configurations M4 and M8 do not provide the appropriate sequence length to apply the test. 

We can apply nine out of fifteen and ten out of fifteen tests for $M32$ and $M64$ configuration, respectively. The list of applicable tests according to sequence length is as follows. The frequency, block frequency, runs, longest runs of ones in a block, serial, approximate entropy, DFT, and cumulative sums, whereas each test has its own significance. For example, two out of these nine tests are used to validate the frequency of 1's and 0's in the sequence, which provides a better assessment of the uniformity.

}

We have produced ROPUF sequences by applying a random initial LFSR seed and test the generated sequences for NIST statistical tests. We have produced a total of $54$ sequences, each from an FPGA device with a single seed value, and out of $54$ at least $51$ sequences (at least 94.44\% sequences) should have to be passed for a confidence interval $\alpha = 0.01$.

\begin{figure}[tbh]
\centering{}\includegraphics[width=1\columnwidth]{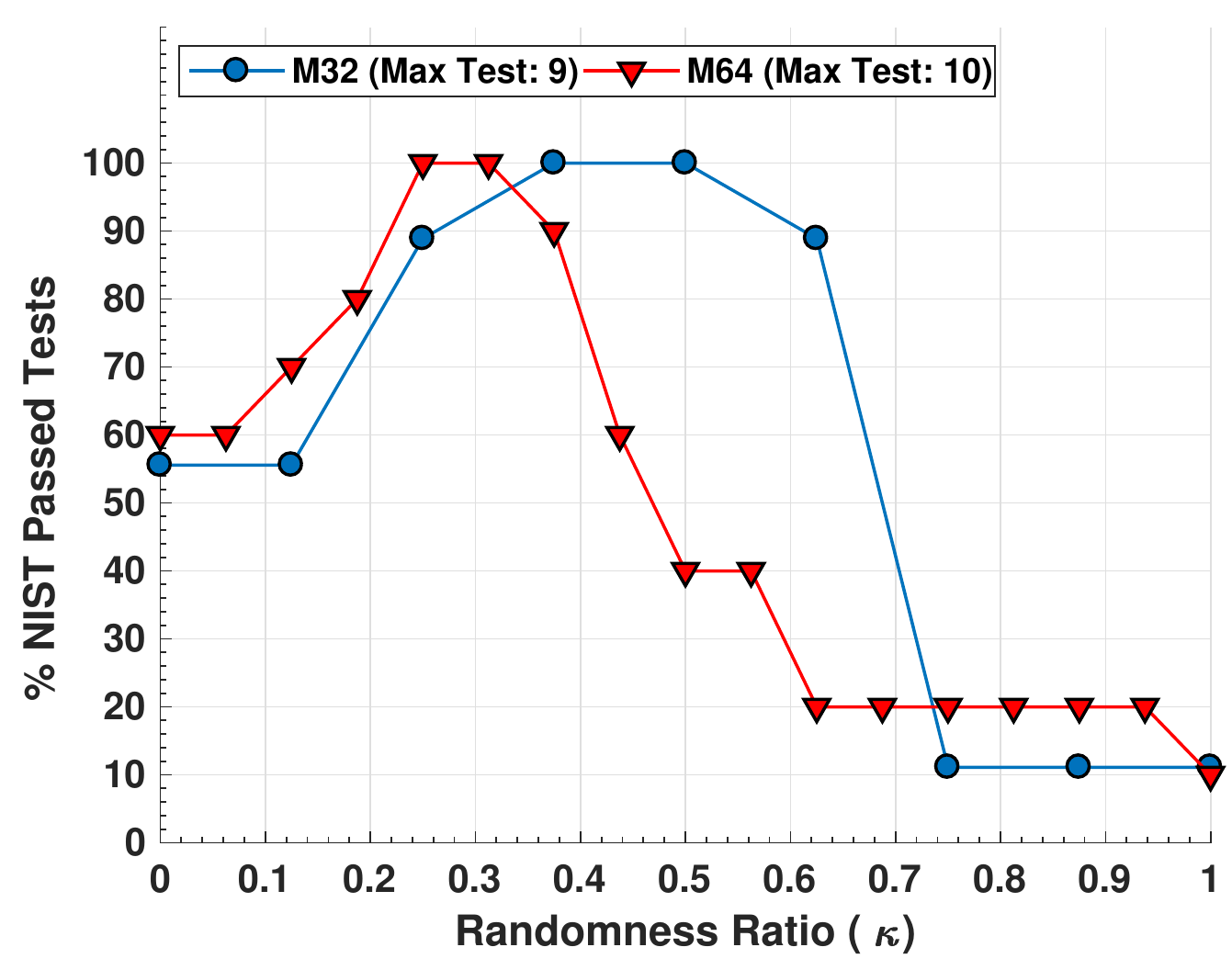}\caption{NIST statistical test pass percentage with the randomness ratio $(\kappa)$ variation. The number of passed tests have $\kappa = 0.375, 0.5$ for M32 and  $\kappa = 0.3125, 0.375$ for M64 configuration.
\label{fig:Kappa_NIST}}
\end{figure}

The effect of randomness factor $\kappa$ has been validated, and the obtained results are shown in Figure \ref{fig:Kappa_NIST}. There are only two possible $\kappa$ values provides 100\% passing rate for $M32$ and two possible $\kappa$ values for $M64$ configurations. For a small value of $\kappa$, frequency and cumulative sum tests have been failed, because of the over-uniform responses. As the $\kappa$ value increases, the uniformity reduces, and randomness increases, which improves the results and further increment in $\kappa$ produces responses with over-randomness, which leads to failed most of the tests. The summarized results for applicable sequences are presented in Table \ref{tab:NIST_table}. There are two possible sequences for M32 and two possible sequences for M64 has passed all the NIST statistical tests.

{\color{blue}

\subsubsection{Minimum Entropy}

The random sequences should have high entropy to pass the NIST statistical test for approximate entropy. Moreover, the entropy extracted through the ring oscillator can be expressed as $log_{2}(M!)$ \cite{Reference_ROPUF_Srinidevdas}. Since, we are using M32 and M64 design, which can provide a maximum of $log_{2}(32!)$ and $log_{2}(64!)$ bit entropy. 

The proposed approach uses a randomized placement and an LFSR design. Therefore, these two techniques certainly improve the minimum bound of the entropy. Whereas, we can measure lower bound of entropy ($\mathcal{H}_{\mathcal{LB}}^i$) for $i^{th}$ bit position and full lower bound of entropy ($\mathcal{H}_{\mathcal{FLB}}$) of the whole sequence using (\ref{eq:min_entropy}) \cite{C_GU_Entropy}.
\begin{equation}
\begin{split}
	\mathcal{H}_{\mathcal{LB}}^i = -log_{2}\left[max(b_{1}^{i},b_{0}^{i})\right]; \\ 
	\mathcal{H}_{\mathcal{FLB}} = \frac{1}{k}\sum_{i=1}^k \mathcal{H}_{\mathcal{LB}}^i ; \\ 
\end{split}
\label{eq:min_entropy}
\end{equation}
Here, $b_{1}^{i}$ and $b_{0}^{i}$ are the number of 1's and 0's in a binary sequence on many hardware for the $i^{th}$ bit location and $k$ is the maximum response bits extracted from single ROPUF instance. The full minimum entropy ($\mathcal{H}_{\mathcal{FLB}}$) for $54$ devices has been evaluated using (\ref{eq:min_entropy}) and depicted in Figure \ref{fig:Kappa_entropy} for both M32 and M64 design with all possible $\kappa$ configurations.

\begin{figure}[tbh]
\centering{}\includegraphics[width=1\columnwidth]{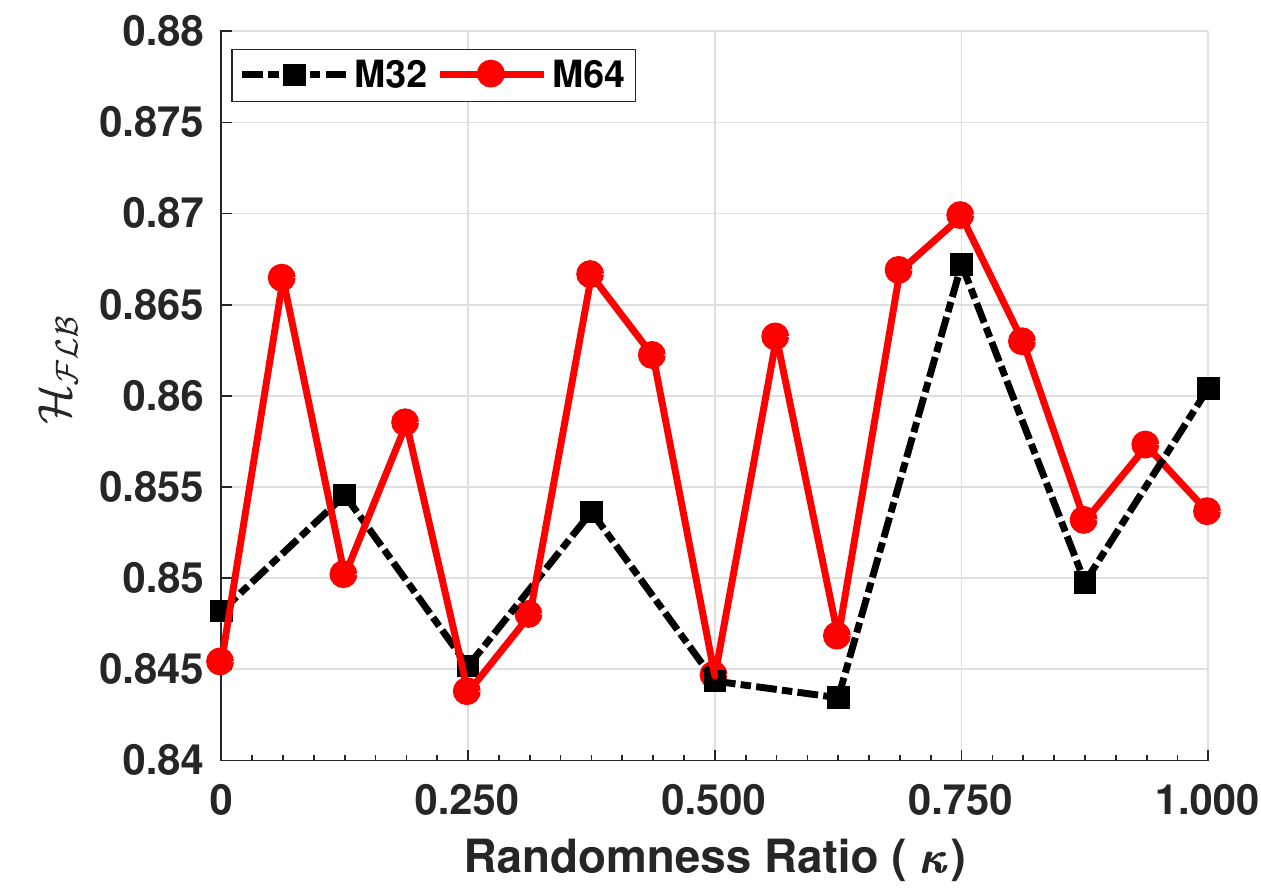}\caption{Full minimum entropy $\mathcal{H}_{\mathcal{FLB}}$ for M32 and M64 with different $\kappa$ configurations}.
\label{fig:Kappa_entropy}
\end{figure}

The mean value of $\mathcal{H}_{\mathcal{FLB}}$ for M32 and M64 are $0.8519$ and $0.8564$, respectively, with all $\kappa$ variation. However, the maximum theoretical value is 1 when the number of 1's and 0's are equally distributed over each bit position. Therefore we have achieved a good minimum entropy value for both M32 and M64 hardware designs.
}

\subsection{Reliability Evaluation}

PUF can produce a random signature from a digital circuit. Moreover, the digital circuit uses the device manufacturing variation, which is uncontrolled and sensitive to environmental noise. Consequently, frequencies of ring oscillator may vary, which produces random bit flipping. The proposed approach increases the minimum pairwise frequency difference ($\chi$) and presuming that it may improve reliability.

\begin{figure}
\centering
\subfloat[][\textbf{VCC Variation }]{\includegraphics[trim = 0cm 0cm 0cm 0cm, width=1 \columnwidth]{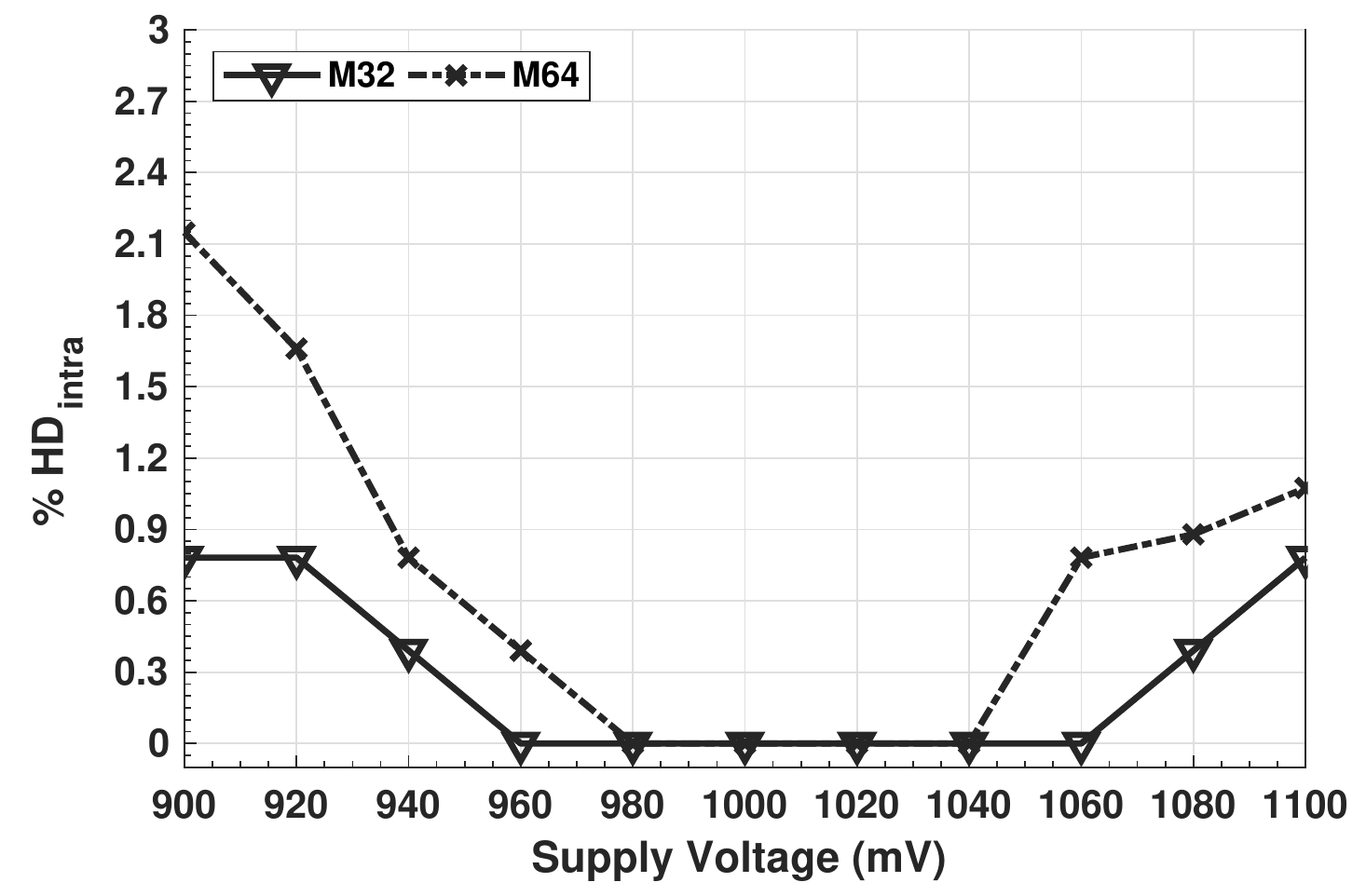} \label{fig:N32_InterHD} }\\
\subfloat[][\textbf{Temperature Variation}]{\includegraphics[trim = 0cm 0cm 0cm 0cm, width=1 \columnwidth]{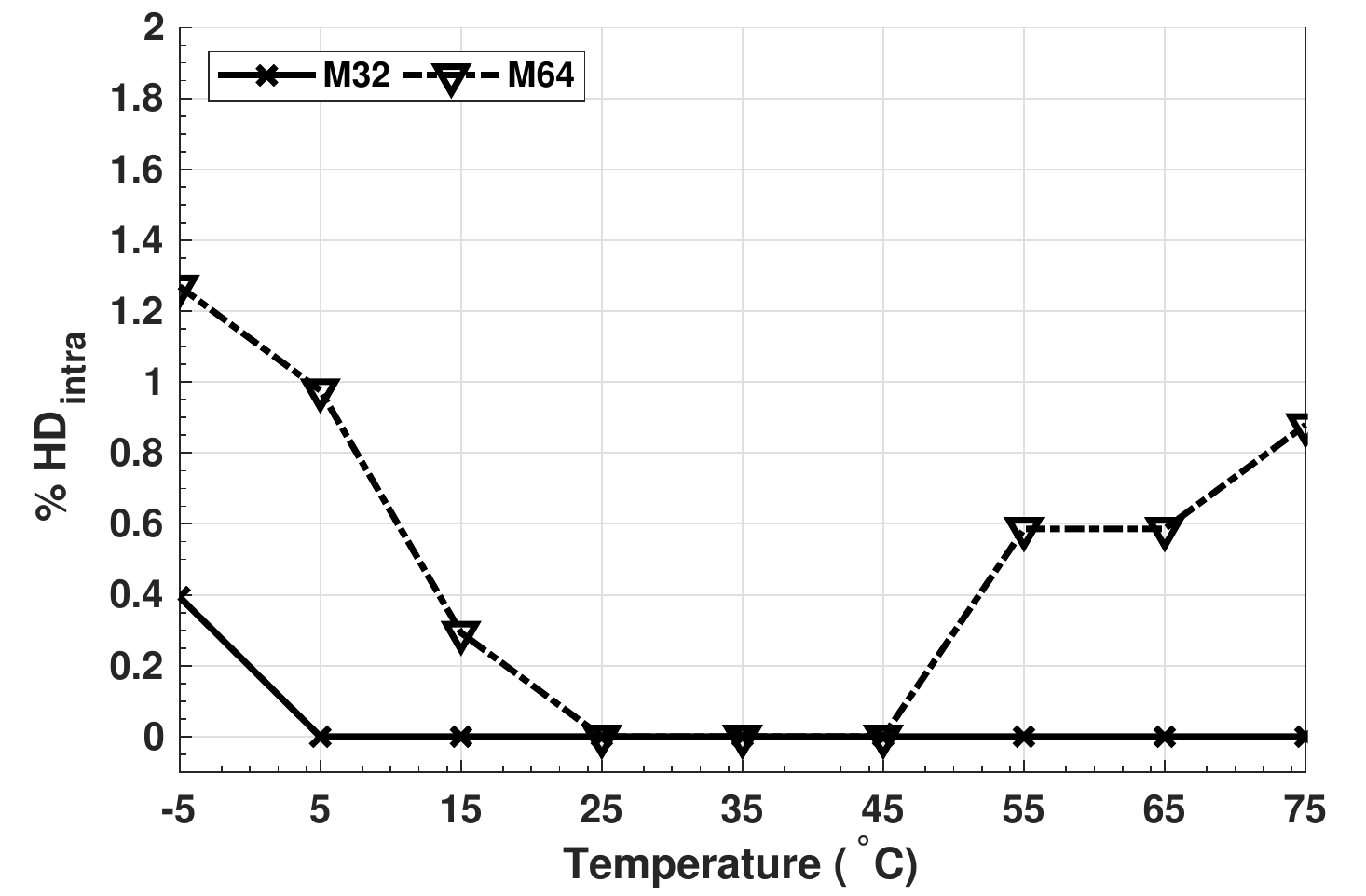}\label{fig:N64_InterHD}}
\caption{\label{fig:INTRA_HD} The effect of environmental variations (temperature and supply voltage) on intra chip hamming distance ($\text{HD}_{intra}$), measured for a single \textit{Basys-3} FPGA device.}
\end{figure}

\begin{figure*}
\centering
\subfloat[][\textbf{$M32$}]{\includegraphics[trim = 0cm 0cm 0cm 0cm, width=1 \columnwidth]{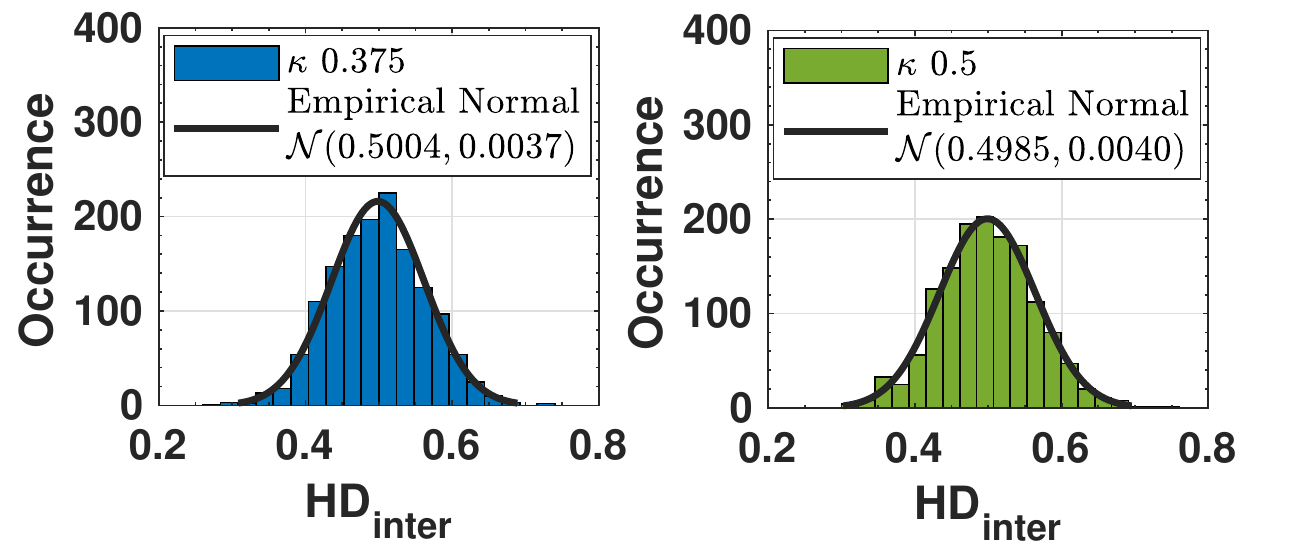} \label{fig:N32_InterHD} }
\subfloat[][\textbf{$M64$}]{\includegraphics[trim = 0cm 0cm 0cm 0cm, width=1 \columnwidth]{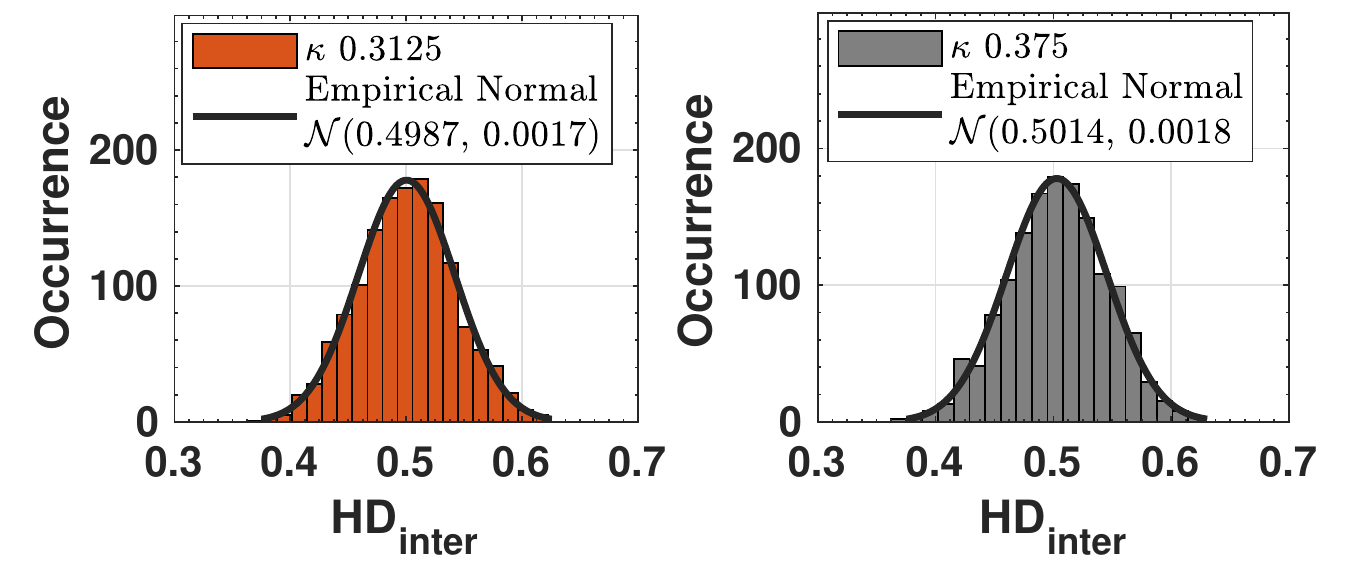}\label{fig:N64_InterHD}}
\caption{\label{fig:INTER_HD} Inter-chip hamming distance ($\text{HD}_{inter}$) for $M32$ with $k = {0.375, 0.5}$ and $M64$ with $k={0.3125,0.375}$. }
\end{figure*}

The reliability is defined for PUF response variation under different environmental conditions $(e)$. The reliability can be evaluated in terms of intra-chip hamming distance ($\text{HD}_{intra}$), which is defined as the number of response ($R$) bit flipping at different environmental conditions out of $k$ bits. The variation has been observed by comparing response ($R_{e}$) with the golden responses ($R_{g}$). The golden response is the response which can be extracted at reference environmental variations. The intra-chip hamming distance for $i^{th}$ PUF can be evaluated using (\ref{eq:intra_HD}).

\begin{equation}
\text{HD}_{intra_{i}} =  \frac{1}{k} \sum_{j=1}^e \textit{HD}(R_{g_{i}},R_{i,j});
\label{eq:intra_HD}
\end{equation}

The reliability metric $r_i$ of $i^{th}$ device can be evaluated using (\ref{eq:Reliability_formula-1}).

\begin{equation}
r_i=\left[ 1-\left( \frac{ \text{HD}_{intra_{i}}}{e} \right) \right] \label{eq:Reliability_formula-1}
\end{equation}

Here, $R_{g_{i}}$ and $R_{i,j}$ both are the $k$ bit response for the $i^{th}$ device. $R_{g_{i}}$ is the reference or golden response and $R_{i,j}$ is the response obtained in the presence of the $j^{th}$ environmental condition.

The proposed improved K-means with centroid relocation enhanced the minimum pairwise frequency difference ($\chi$), therefore reliability improvement is apparent. The evaluation metric $\text{HD}_{intra}$ in the presence of supply voltage and temperature variation is manifested in Figure \ref{fig:INTRA_HD}. The worst case reliability for $M32$ and $M64$ have been improved by 1.18\% and 0.78\% in the presence of voltage variation. 

The sequences obtained through $M8$ and $M16$ provide a response length of $15$ and $63$ bits, therefore the reliability for these approaches are far better than $M32$. The increment in the $M$ value can reduce minimum pairwise frequency difference $(\chi)$, consequently, the reliability has been reduced. Therefore, we are considering the large value of M, i.e. $M32$ and $M64$, which have already passed NIST statistical test, and assumed that reliability for $M8$ and $M16$ is better than $M32$ as well as $M64$.

The average reliability obtained through the Figure \ref{fig:INTRA_HD} has been summarized in Table \ref{tab:relability_variation}. The average reliability improvement for M32 and M64 are 0.35\% and 1.14\%, respectively.

\begin{table}[]
\renewcommand{\arraystretch}{1.2}
\small
\caption{\% The summarized reliability results in the presence of environmental variations. Here, $r_{max}$, $r_{min}$ and $r_{avg}$ are the maximum, minimum and average reliability values}
\label{tab:relability_variation}
\begin{tabular}{|l|l|l|l|l|l|l|}
\hline
\multirow{2}{*}{\textbf{M}} & \multicolumn{3}{c|}{\textbf{VCC Variation}}                                                                                 & \multicolumn{3}{c|}{\textbf{Temp Variation}}                                                           \\ \cline{2-7} 
                            & \multicolumn{1}{c|}{\textbf{$r_{max}$}} & \multicolumn{1}{c|}{\textbf{$r_{min}$}} & \multicolumn{1}{c|}{\textbf{$r_{avg}$}} & \multicolumn{1}{c|}{\textbf{$r_{max}$}} & \multicolumn{1}{c|}{\textbf{$r_{min}$}} & \textbf{$r_{avg}$} \\ \hline
\textbf{32}                 & 1                                       & 0.9921                                  & 0.9961                                   & 1                                       & 0.9961                                  & 0.9980              \\ \hline
\textbf{64}                 & 1                                       & 0.9785                                  & 0.9890                                   & 1                                       & 0.9873                                  & 0.9936             \\ \hline
\end{tabular}
\end{table}

\begin{center}
\begin{table*}
      \centering
      \renewcommand{\arraystretch}{2}
\small
      \caption{Comparison of the group of  PUF characteristics of the proposed scheme with those of earlier schemes}
      \label{tab:state_of_the_art}
      \centering
  {\color{blue}

		\begin{tabular}{|c|c|c|c|c|c|c|c|c|c|c|c|}
		
\hline
\multirow{2}{*}{Method} & \multirow{2}{*}{\textbf{\%$r_{avg}$}} & \multirow{2}{*}{\textbf{\%$u$}} & \multirow{2}{*}{\textbf{$k$}} & \multirow{2}{*}{\textbf{Resources}} & \multicolumn{2}{c|}{\textbf{Resources/bit}} & \multirow{2}{*}{\textbf{FPGA}} & \multirow{2}{*}{\textbf{Tech.}} & \multirow{2}{*}{\textbf{$\Delta T$}} & \multirow{2}{*}{\textbf{$\frac{\Delta V}{V_{ref}}$}} &\multirow{2}{*}{\begin{tabular}[c]{@{}c@{}}NIST\\ ($\frac{App.}{Total}$)\end{tabular}} \\ \cline{7-7}
                        &                       &                    &                    &                           &                 &   \textbf{Norm}            &                       &                       &                          &                          &                      \\
\hline
ROPUF \cite{Reference_ROPUF_Srinidevdas} $^\star$                                                                                 & 99.52                  & 46.15                 & 128                & 1024 ROs &    8.0000  & 63.79     & Vertex-4 & $90$nm           &                                                                  \textbf{100}$^\circ C$ & $20\%$ & \textbf{?}                                                                     \\ \hline 
Maiti-CRO \cite{Maiti_J_Crypto}                  & 99.14                  & 47.31                 & 127                & 512 Slices   & 4.0314 & 32.14 & Spartan-3E  & $90nm$        &                                                                  $40 ^\circ C$ & \textbf{40}\% &  \textbf{?}                                                                    \\ \hline 

Improved ROPUF \cite{NIST_Reviewer_1}  &  95\% &  49.32  & 150 & 1200 Slices & 8.0000 & 63.79 & Spartan3E & 90nm  &  \textbf{?}   & 0.83\%  &  $\frac{8}{15}$ {\color{black}\cmark}                                                                   \\ \hline

Self Compare \cite{Selfcompare}    $^\dagger$             &        \textbf{99.80}           & 49.07                 & \textbf{256}                & \textbf{4 LUTs}  &    \textbf{0.0039}   & 0.032     & Spartan-6  & $45$nm          &                                                                $65^{\circ} C$   & \textbf{?} &   \textbf{?}                                                                  \\ \hline 
Compact RO \cite{Low_Area_Reconfigurable_PUFs}                 & 99.16                  & 47.13                 & \textbf{256}                & 32 Slices   &   0.1250 & 0.996    & Spartan-6  & $45$nm          &                                                                 \textbf{?}  & \textbf{?} &   {\color{red}\xmark}                                                                     \\ \hline 
PUF-ID   \cite{PUFID_Generator_GU_ET_AL} $^\ddagger$                 & $\mathrm{99.40}$                      &  45.60                     &    128                &      128 Slices    &  1.0000 & 7.974    & \textbf{Artix-7} & \textbf{28nm}         & $75 ^\circ C$                                                                  & 20\%  &    \textbf{?}                                                                      \\ \hline 
Biased ROPUF \cite{Arjun_IEEE_CONECCT}             & 99.05                  & 35.83                 & \textbf{256}                & 32 Slices          & 0.1250  & 0.996 & \textbf{Artix-7} & \textbf{28nm}            &                                                                  $40 ^\circ C$ &             \textbf{?}   &   {\color{red}\xmark}                                                        \\ \hline 
Previous Work \cite{ARJUN_VLSID}             & 99.35                  & 49.83                 & 255                & 32 Slices     & 0.1254 & 1.000      & \textbf{Artix-7} & \textbf{28nm}            &      80$^\circ C$                                                             &         20\%  &  $\frac{9}{15}$  {\color{black}\cmark}                                          \\ \hline 

Proposed Work (M32)              & 99.70                  & \textbf{49.90}                 & 255                & 32 Slices          & 0.1254 & 1.000 & \textbf{Artix-7} & \textbf{28nm}            &      80$^\circ C$                                                             &         20\%    &   $\frac{9}{15}$ {\color{black}\cmark}                                        \\ \hline

 \hline                 
\end{tabular}
\begin{tablenotes}
      \footnotesize
      \item \textbf{?} = Failed/Not reported. \hspace{2em} $k$ = Maximum Response Length. \hspace{2em} $\Delta T = T_{max} - T_{min}$ (Temperature Variation Range). 
      \item $\Delta V = V_{max} - V_{min}$ (Supply Voltage Variation Range). \hspace{2em} $^\dagger$ = Required resources excluding adaptive tuning circuits. 
      \item  $^\star$ = Assuming that 1-RO requires 1 Slice. \hspace{2em} $^\ddagger$ = Post processing.  \hspace{2em} $V_{ref}$ = Reference voltage. \hspace{2em} \textbf{Norm}= Normalized Resources/bit.
      \item NIST =  NIST test status. ($\frac{\text{Applicable}}{\text{Total}}$) and ({\color{black}\cmark} = Passed, {\color{red}\xmark} = Failed)
    \end{tablenotes}
    }
\end{table*}
\end{center}

\subsection{Uniqueness}
The uniqueness is quite an important metric to ensure that two devices are not producing the same responses for the same applied challenges. The uniqueness can be evaluated using inter-chip hamming distance ($\text{HD}_{inter}$). The inter-chip hamming distance is evaluated in between $q$ number of different devices. The inter-chip variation appears due to the inter-chip random variation. The inter-chip hamming distance ($\text{HD}_{inter}$) in between $q$ number of different chip can be evaluated using (\ref{eq:Inter_HD}).

\begin{equation}
\text{HD}_{inter}=\frac{1}{k}\sum_{i=1}^{q-1} \sum_{j=i+1}^{q}  \text{HD}(R_{i},R_{j}) 
\label{eq:Inter_HD}
\end{equation}

Here, $R_{i}$ and $R_{j}$ are the $k$ bit responses and $i\neq j$. The theoretical value of the interchip hamming distance is $\text{HD}_{inter} = 0.5$. The uniqueness metric $(u)$ is evaluated using (\ref{eq:uniquness_form}).

\begin{equation}
u=\frac{2\times \textit{HD}_{inter}}{q (q-1)}  \label{eq:uniquness_form}
\end{equation}

Figure \ref{fig:INTER_HD} represents the inter chip hamming distance $(\text{HD}_{inter})$ for $M32$ and $M64$ configuration. Figure \ref{fig:N32_InterHD} and \ref{fig:N64_InterHD} represents the inter-chip hamming distance for M32 with $\kappa = 0.375, 0.5$ and M64 with $\kappa = 0.3125, 0.375$. The distribution obtained through inter-chip hamming distance looks like Gaussian distribution, therefore we have fit the distribution with standard normal distribution and empirically determined the statistical parameters for inter-chip hamming distance. The mean value for each distribution is near to theoretical value $0.5$. 

The randomized placement deforms the sorting order of the placement locations, therefore the output response is random, and the responses obtained from the different devices are randomized for the same challenge ($C$). The improvement in the uniqueness is significant, and the uniqueness value is near to theoretical value. The obtained average uniqueness value has a maximum deviation of $0.22\%$ as compared to the theoretical value.

\subsection{Comparison}

{\color{blue}
The proposed method is examined along with some of the existing approaches, and a comparison has been presented in Table \ref{tab:state_of_the_art}. The reliability metric is the quite essential in order to evaluate the reproducibility of the sequences because the non-reproducible sequences can not be utilized for device authentication application. Moreover, the standalone ROPUF can not defend against ML attacks and modeling attacks.  

All the existing approaches in the group have reliability $>99\%$ except \cite{NIST_Reviewer_1}, which is not considered for reliability comparison. The approach \cite{Selfcompare} has the highest reliability, whereas our method occupies the second position among the group. However, this approach \cite{Selfcompare} doesn't take voltage variations into consideration.  We have considered both supply voltage and the temperature variation, and even the temperature range is higher by $15^\circ C$ compared to the method in \cite{Selfcompare}. Moreover, our approach provides 100\% reliability in the same temperature variation range, as mentioned in \cite{Selfcompare}. The method in \cite{PUFID_Generator_GU_ET_AL} provides reliable responses in the presence of both temperature and supply voltage variation, however, it requires post-processing to achieve that much reliability. The approaches \cite{Low_Area_Reconfigurable_PUFs} \cite{Arjun_IEEE_CONECCT}, provide reliable response with no environmental variations. The approaches \cite{Maiti_J_Crypto} \cite{Reference_ROPUF_Srinidevdas}, provide better reliability as compared to all other approaches, except our approach, in the presence of environmental variations. Therefore if we are considering the effect of environmental variations, then our approach shows superiority among the considered approaches in the group.

Our proposed approach requires $32$ slices in order to produce $255$ bit CRPs. The method in \cite{Selfcompare} provide minimum utilization of the slices, however, the area for adaptive tuning circuit on hardware requires extra circuitry, which is not reported in the paper. The methods  \cite{Low_Area_Reconfigurable_PUFs} \cite{Arjun_IEEE_CONECCT} \cite{ARJUN_VLSID} occupy similar number of slices in comparison to our present approach. Furthermore, the resources required to produce a single bit response is approximately the same as reported in \cite{Arjun_IEEE_CONECCT} \cite{ARJUN_VLSID}.

The approach in \cite{Reference_ROPUF_Srinidevdas}, provides significantly good reliability with the comparatively large environmental variation range. However, the method consumes $32$ times extra slices to implement ring oscillators in order to produce the same number of response bits. The approach in \cite{Maiti_J_Crypto}, provides a reliable response, but the uniqueness is less and area utilization is significantly large i.e. approximately 30 times.

The randomized placement provides the highest uniqueness value in the group of existing PUF design. The method in \cite{Selfcompare} and \cite{NIST_Reviewer_1} provide a uniqueness value $>49\%$. However, \cite{Selfcompare}  requires extra computation time and resource utilization for adaptive tuning, and the method in \cite{NIST_Reviewer_1} has poor reliability. Apart from \cite{Selfcompare} and \cite{NIST_Reviewer_1}, no other approach is able to achieve that much of uniqueness value. 

Finally, one of the strengths of the present approach is \textit{randomness}. The proposed approach passes all nine applicable NIST statistical test by employing random group allocation of the ring oscillator. We have included a configurable randomness factor $\kappa$, which allows passing all nine tests for two randomness factor. The approaches in \cite{NIST_Reviewer_1} passed eight tests out of nine applicable tests. Apart from these, no other method in the group is bale to pass the NIST statistical tests.

In summary, the proposed approach has these features, which show the superiority over the existing approaches- i) excellent reliability despite environmental variations, ii) enhanced uniqueness, iii) improved randomness, iv) hardware/area efficient and v) implemented on recent technology (28nm). All these features are simultaneously available in the cost of one additional phase for characterization, which can be accomplished with auto characterization framework at the time of PUF creation.
}

\subsection{Security Threats}

As the proposed approach is engaged in two phases, therefore, the information leakage during PUF creation can produce serious threats. Moreover, the PUF creation environment should be secure. If an attacker is able to extract the frequency information, then it is possible for him to evaluate centroid frequency by employing the proposed approach. The randomized placement reduced the chances to predict responses bits because the frequency ordering and the placement is random. {\color{blue} Moreover, the ROPUF architecture is the same as earlier; therefore, it can not resists against the modeling attacks, i.e. GA based modeling \cite{GA_Based_Attack} and quicksort based modeling \cite{ML_Attacks}. However, the proposed approach enhanced the performance of all three essential features such as reliability, uniqueness and randomness. Therefore it can be used with some other true random number generator (TRNG) or pseudo-random number generator (PRNG) to improve security performance.}

%\newpage
\section{Conclusion\label{sec:Conclusions}}

{\color{blue}
The proposed method using novel augmentations provide significant improvement in reliability as well as uniqueness, simultaneously despite voltage as well as temperature variations. The proposed approach has been primarily implemented and validated on \textit{28nm-technology Xilinx} FPGA. The randomness for the proposed method passes NIST statistical tests for randomness facilitated by randomized placement. The approach requires two-phases- the first phase requires  one-time additional efforts (trusted environment) to perform frequency pre-characterization on FPGA; whereas the second phase is utilized for RO selection in order to maximize frequency difference. 
The ring oscillator PUF design proposed in this paper can be used along with with other circuitry like TRNG/PRNG to improve the machine learning resistivity. The work on device authentication protocol and the reliability examination against device aging is currently being pursued by us.

}

\begin{acknowledgements} 
{
We also gratefully acknowledge SMDP-C2SD project (ODRC No. 1000110086) project funded by \textit{Ministry of Electronics \& IT, Government of India} for technical support. \vspace{0.25em}

We are thankful to Mr Bharat Kasyap, MNIT, Jaipur to provide support for experimental setup design.
}
\end{acknowledgements}

% Authors must disclose all relationships or interests that 
% could have direct or potential influence or impart bias on 
% the work: 
%
% \section*{Conflict of interest}
%
% The authors declare that they have no conflict of interest.

\Urlmuskip=0mu plus 1mu\relax
% BibTeX users please use one of
%\bibliographystyle{spbasic}      % basic style, author-year citations
\bibliographystyle{spmpsci}      % mathematics and physical sciences
\balance
\bibliography{Springer_Reference_JETTA_Review_1_ver2}   % name your BibTeX data base

%% Non-BibTeX users please use
%\begin{thebibliography}{}
%%
%% and use \bibitem to create references. Consult the Instructions
%% for authors for reference list style.
%%
%\bibitem{Reference_JETTA.bib}
%% Format for Journal Reference
%Author, Article title, Journal, Volume, page numbers (year)
%% Format for books
%\bibitem{RefB}
%Author, Book title, page numbers. Publisher, place (year)
%% etc
%\end{thebibliography}

\end{document}